# Phase-field simulations of nucleation, growth, and coarsening of $\beta_1$ precipitates in Mg–Nd alloys

L. Shi[1], S. DeWitt[1], D. Montiel[1], Q. Shi[1], J. Allison[1], and K. Thornton[1,2,*]

[1]Department of Materials Science and Engineering, University of Michigan, Ann Arbor, MI 48109, United States of America

[2]Department of Nuclear Engineering and Radiological Sciences, University of Michigan, Ann Arbor, MI 48109, United States of America

*Author to whom any correspondence should be addressed.

**Email**: kthorn@umich.edu

## Abstract

The spatial distribution and morphology of precipitates formed during aging are key factors that determine the precipitation-hardening response of various magnesium–rare-earth alloys. In recent years, the use of high-performance computing clusters and massively parallel frameworks has enabled quantitative simulations of the evolution of individual and multiple precipitates at relevant length and time scales. However, predictive modeling of precipitate evolution remains challenging, in part because many key thermodynamic and kinetic parameters governing the underlying physics are either unknown or have a high degree of uncertainty. In this work, we developed a workflow in which experimental data were used to parameterize a phase-field model to perform two-dimensional (2D) simulations of concurrent nucleation and evolution of $\beta_1$ precipitates in magnesium–neodymium alloy during aging. Matrix composition and precipitate number density at different aging times were obtained from atom probe tomography and transmission electron microscopy measurements, respectively. We applied a stereological method to estimate the three-dimensional (3D) number densities from experimental cross-sectional transmission electron micrographs. The estimated 3D number density data were then converted to effective 2D number densities. The effective 2D number density and composition data were used to determine the required model parameters by minimizing the discrepancy between simulation and experimental results. The parameterized model enables quantitative phase-field simulations of nucleation and growth of $\beta_1$ precipitates, which can be employed to optimize aging time to achieve a target number density of precipitates. This work demonstrates a strategy to overcome the challenges associated with parameterizing a coupled phase-field and nucleation model.

## 1. Introduction

Magnesium (Mg) alloys hold significant promise as structural materials in the automotive and aerospace industries due to their high strength and low density [1-3]. The demand for lightweight materials is increasing, as they improve fuel efficiency and reduce emissions by enabling significant weight savings in automobiles [4]. The incorporation of rare-earth (RE) elements into Mg alloys enhances their mechanical strength and creep resistance [5]. Furthermore, Mg–RE alloys offer excellent ignition resistance [6] and, in some cases, enhanced corrosion resistance [7], while maintaining low density [2, 8, 9]. The enhanced mechanical properties are largely due to precipitation strengthening, in which plate-like precipitates with habit planes perpendicular to the basal plane form during aging [3, 10, 11].

Understanding the mechanisms that govern the formation and growth of precipitates is crucial for designing Mg–RE alloys with tailored thermomechanical properties. Notably, among the RE elements, neodymium (Nd) exhibits the most pronounced age hardening when alloyed with Mg. This is primarily attributed to the high solid solubility of Nd in Mg, which allows for a high concentration of Nd during solution treatment [12, 13] and the subsequent formation of fine precipitate phases (e.g., $Mg_3Nd$) during aging, leading to a significant increase in hardness and strength. Consequently, Mg–Nd and Mg–Y–Nd alloys (with Y representing yttrium), which benefit from conventional age-hardening processes, are extensively used in aerospace and transportation applications [14, 15].

Experimental and computational studies have been conducted to investigate the phases, morphologies, and spatial arrangements of precipitates formed during the aging of Mg–Nd alloys. Natarajan et al. [8] performed a combined first-principles and experimental study to elucidate the precipitation sequence in binary Mg–Nd alloys. This sequence typically proceeds from a supersaturated solid solution (SSSS) to Guinier-Preston (GP) zones, followed by the precipitate phases $\beta'''$, $\beta_1$, $\beta$, and $\beta_e$ in that order. Ji et al. [16] utilized first-principles calculations and phase-field simulations to predict $\beta'$ precipitate morphology. Liu et al. [17] used phase-field modeling to investigate the morphology of $\beta_1$ precipitates in Mg–Nd alloys. DeWitt et al. [18] examined the composition and morphology of coherent $\beta'''$ precipitates, comparing predictions from three-dimensional (3D) phase-field simulations with high-angle annular dark field (HAADF) scanning transmission electron microscopy (STEM) measurements. Zhu et al. [19] employed atomic-resolution HAADF–STEM to identify impinged $\beta_1$ variants in a characteristic triadic configuration.

While most studies have provided significant insights into the phases, morphologies, and spatial arrangements of precipitates in Mg–Nd alloys, less attention has been given to the evolution of precipitate populations across the stages of nucleation, growth, and coarsening. Coarsening is the process in which larger precipitates grow at the expense of smaller ones, driven by the reduction of interfacial energy [20, 21]. Understanding how precipitate number density changes during aging helps in choosing optimal time and temperature conditions to achieve a precipitate number density that results in the desired mechanical properties in Mg–Nd alloys. However, in many phase-field modeling studies, including but not limited to Mg–Nd systems, the evolution of precipitate number density across all aging stages is rarely examined. For example, Yang et al. [22] employed a phase-field model to investigate the effect of alloy composition on the evolution of precipitate number density, but they did not consider the coarsening stage. Liu et al. [23] focused on how number density affects alloy strengthening, without tracking its evolution across all aging stages. Ji et al. [24] examined morphological changes and interactions among precipitates rather than changes in precipitate number density.

A challenge in comparing the precipitate number densities from phase-field simulations and experimental characterization, e.g., transmission electron microscopy (TEM), is that experimental measurements are typically obtained from two-dimensional (2D) cross-sections, which may not accurately reflect the number density within the full 3D microstructure. On the simulation side, while 3D phase-field simulations provide a more realistic representation of the microstructure, simulating a statistically representative 3D sample often requires a significant amount of computational resources in general. As further discussed in section 4.4, for systems like Mg–Nd alloys, the computational cost for phase-field simulations is especially high because the precipitates exhibit strong anisotropy, and a high spatial resolution is required to accurately describe their morphology. Therefore, performing 2D phase-field simulations is often a practical choice. However, this approach requires careful stereological conversion strategies to parameterize the model, so that the number density evolution observed in 2D simulations represents the 3D evolution as closely as possible.

In this work, an integrated experimental and phase-field workflow was developed to model the evolution of $\beta_1$ precipitates in Mg–Nd alloys during aging, which involves nucleation, growth, and coarsening. Our proposed workflow involves extracting 3D precipitate number densities from 2D experimental images and using these data to obtain input parameters for 2D phase-field simulations. The phase-field model implementation includes a stochastic method to introduce nuclei during the simulation. The model was parameterized using a combination of available literature data and experimental measurements, including matrix composition obtained via atom probe tomography (APT) and number density extracted from TEM images at various aging times. A validated model through this approach enables the prediction of process-structure relationships that allow for the optimization of aging conditions to achieve desired microstructures.

This paper is organized as follows. Section 2 describes the experimental procedures for measuring precipitate number density and average matrix composition at different aging times. Section 3 presents the simulation methods. Section 4 presents the results, including the experimental measurements, the model parameterization workflow and outcomes, the simulation results, and the comparison between the simulation and experimental results. We find that the parameterized model is capable of capturing the salient features of nucleation, growth, and coarsening behaviors observed in experiments. Section 5 summarizes the main conclusions of this work.

## 2. Experimental methods

### 2.1. Sample preparation

As-cast Mg–2.4 wt.% Nd billets, purchased from CanmetMATERIALS, Canada, were used in this study. The purchased billets were melted by resistance heating in a steel crucible, under a protective gas mixture of $CO_2$ and 0.5% $SF_6$ in an electric furnace, which was then poured into a mold to form ingots with a diameter of 3.5 inches. To produce $\beta_1$ precipitates, the cast alloy was first homogenized at 803 K for 3 h. The homogenized material was rapidly quenched in water and subsequently aged at 523 K to facilitate the precipitation of $\beta_1$ phase in Mg–Nd alloy. Further details regarding the casting, homogenization, quenching, and aging procedures can be found in Refs. [25, 26]. A total of five aging times were investigated: 3 h, 3.25 h, 4 h, 6 h, and 9 h at 523 K.

## 2.2. Determination of precipitate number density

TEM specimens were mechanically thinned to 100 μm and electropolished using a twin-jet polisher. High-resolution bright-field TEM images were obtained in STEM mode with a double aberration-corrected JEOL 3100R05 TEM operated at 300 kV. More details of this procedure can be found in Refs. [25, 26]. The TEM images acquired at different aging times were analyzed to determine the number densities of $\beta_1$ precipitates. However, the number of precipitates obtained by direct counting in a TEM image corresponds only to those present within the thin foil. Therefore, the precipitate count in the TEM image cannot be used to directly calculate the number density in the bulk material. To estimate the 3D number densities of precipitates from these 2D TEM images, a stereological correction was necessary. Several stereological methods have been reported for converting 2D measurements obtained from thin sections to 3D number densities, and these often require knowledge of the particle size distribution [27, 28]. Since the precipitate size distribution was not known in our case, we adopted the correction method described by Sonderegger [29], which does not rely on knowledge of the precipitate size distribution. Although this method was originally developed for spherical particles, it is straightforward to show that it can also be applied to plate-shaped particles with a circular cross-section perpendicular to the imaging plane, as shown in section S1 of the Supplementary Information (SI).

## 2.3. Determination of matrix composition via APT

The matrix composition at different aging times was quantified via APT. Focused ion beam (FIB) lift-out was performed on a FEI Helios 650 dual-beam system to prepare needle-shaped APT specimens. A Cameca LEAP 5000XR instrument, operated in laser-pulsing mode with a pulse energy of 40–60 pJ and a specimen temperature of 30 K at a detection rate of 1.0%, was used for the APT data acquisition. Data visualization and analysis were performed using the Integrated Visualization and Analysis Software (IVAS®) package. 3D tomographic reconstruction was performed with the radius evolution at shank mode, in which the shank angle was estimated from the high-resolution scanning electron microscopy (SEM) image of the APT needles. The following parameters were employed: voxel size = $1\times1\times1$ nm$^3$, delocalization distance = 3 nm, and sample count threshold = 1%.

# 3. Simulation methods

## 3.1. Phase-field model

During the transformation from the α-Mg phase with a hexagonal close-packed (hcp) structure to the $\beta_1$ phase with a face-centered cubic (fcc) ordering, which involves deformation and rigid body rotation, the symmetries of the α-Mg and $\beta_1$ lattices allow for three possible lattice correspondences [15]. As shown in figure 1, each correspondence variant (assigned red, blue, or green) has two rotation variants [15], generated by rotating the deformed lattice clockwise or counterclockwise by 5.3°. The resulting lattice deformations between $\beta_1$ precipitates (schematic atoms drawn in dark red) and the α-Mg matrix (schematic atoms drawn in dark blue) are noted in figure 1. The two rotation variants of the same correspondence variant have the same stress-free transformation strain (SFTS), and thus we use a single order parameter to represent them, i.e., each correspondence variant, $p$, is represented by $\eta_p = 1$ and $\eta_{q \neq p} = 0$, where $p, q = 1,2,3$. The hcp α-Mg matrix phase is represented by $\eta_1 = \eta_2 = \eta_3 = 0$. Furthermore, each corresponding variant has two primary growth directions within the basal plane, indicated by schematic precipitate shapes in figure 1. This results in microstructures consisting of $\beta_1$ plate-shaped precipitates oriented either nearly parallel or at approximately 120° with respect to $[11\bar{2}0]_\alpha$.

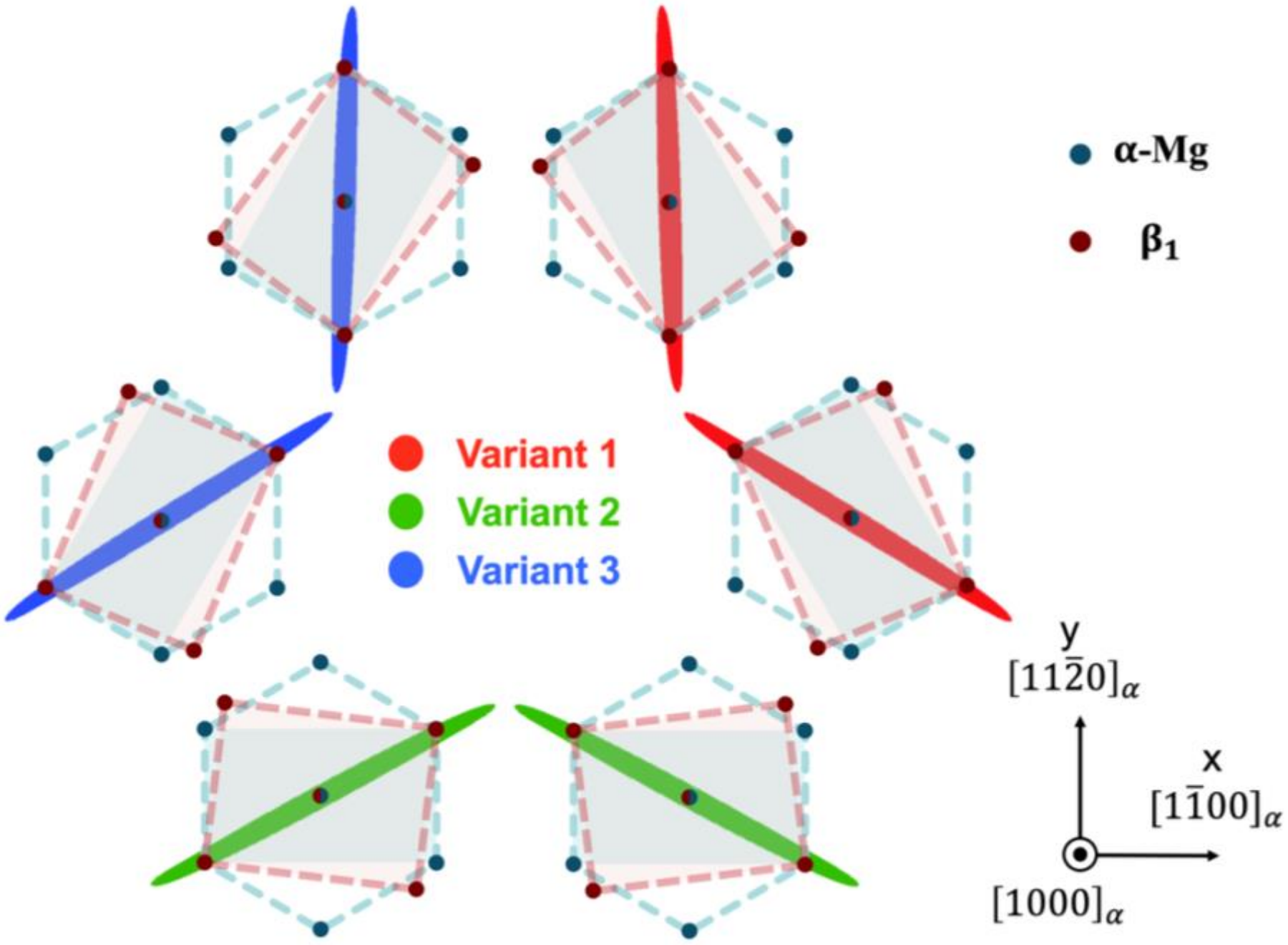

**Figure 1.** Schematic diagram illustrating lattice deformation between $\beta_1$ precipitates (pink dashed outlines with dark red dots indicating the lattice sites) and α-Mg matrix (light blue dashed outlines with dark blue dots indicating the lattice sites) in the basal plane of the α-Mg matrix. The possible primary growth directions of $\beta_1$ precipitates within the basal plane are indicated schematically with representative precipitate shapes, with three correspondence variants in three different colors.

An additional variable, $x_{Nd}$, is used in the phase-field model to represent the Nd composition in mole fraction. In this work, we employed the Kim–Kim–Suzuki (KKS) model [17, 30], following the specific formulation by DeWitt et al. [18]. The total free energy functional is expressed as

$$F_{tot} = \int \left[ f_{bulk}(x_{Nd}, \eta_1, \eta_2, \eta_3) + f_{grad}(\nabla\eta_1, \nabla\eta_2, \nabla\,\eta_3) + f_{el}(\eta_1, \eta_2, \eta_3, \boldsymbol{u}) \right] dV, \quad (1)$$

where $f_{bulk}(x_{Nd}, \eta_1, \eta_2, \eta_3)$ represents the bulk free energy density, $f_{grad}(\nabla\eta_1, \nabla\eta_2, \nabla\,\eta_3)$ represents the gradient energy density, and $f_{el}(x_{Nd}, \eta_1, \eta_2, \eta_3, \boldsymbol{u})$ is the elastic strain energy density. The dependence of the elastic strain density on the displacement vector, $\boldsymbol{u}$, will be discussed later in the text.

The bulk free energy density, which is a function of the composition and three order parameters, is expressed as [18]

$$f_{bulk}(x_{Nd},\eta_1,\eta_2,\eta_3) = f_\alpha(x_{Nd}^\alpha)\cdot\left(1-\sum_{p=1}^{3}H(\eta_p)\right) + f_{\beta_1}\left(x_{Nd}^{\beta_1}\right)\cdot\sum_{p=1}^{3}H(\eta_p) + f_{barrier}(\eta_1,\eta_2,\eta_3). \quad (2)$$

Here, $x_{Nd}^\alpha$ and $x_{Nd}^{\beta_1}$ are the auxiliary Nd compositions for the α-Mg and $\beta_1$ phases in the KKS model, respectively, and $f_\alpha(x_{Nd}^\alpha)$ and $f_{\beta_1}\left(x_{Nd}^{\beta_1}\right)$ are the chemical free energy densities of α-Mg and $\beta_1$ phases, respectively. The constraint $x_{Nd} + x_{Mg} = 1$ is assumed for each phase, where $x_{Mg}$ is the composition of Mg. The polynomial $H(\eta_p) = {\eta_p}^3(10 - 15\eta_p + 6{\eta_p}^2)$ interpolates the bulk free energy densities between the α-Mg and $\beta_1$ phases. The term $f_{barrier}(\eta_1,\eta_2,\eta_3)$, defined below, contributes to the precipitate-matrix interfacial energy and penalizes overlaps of different correspondence variants.

Polynomial forms of $f_\alpha(x_{Nd}^\alpha)$ and $f_{\beta_1}\left(x_{Nd}^{\beta_1}\right)$ are used in this model, as described in Ref. [18]:

$$f_\alpha(x_{Nd}^\alpha) = A_0 + A_1 x_{Nd}^\alpha + A_2 (x_{Nd}^\alpha)^2, \quad (3)$$

$$f_{\beta_1}\left(x_{Nd}^{\beta_1}\right) = B_0 + B_1 x_{Nd}^{\beta_1} + B_2\left(x_{Nd}^{\beta_1}\right)^2. \quad (4)$$

Here, $A_0$, $A_1$, $A_2$, $B_0$, $B_1$ and $B_2$ are constants defining the chemical free energy density functions. These constants were obtained by performing quadratic fits to the chemical free energy densities calculated using the CALPHAD (CALculation of PHAse Diagrams) method near equilibrium compositions. The auxiliary compositions, $x_{Nd}^\alpha$ and $x_{Nd}^{\beta_1}$, are given by [18]

$$x_{Nd}^\alpha(x_{Nd},\eta_1,\eta_2,\eta_3) = \frac{B_2 x_{Nd} + \frac{1}{2}(B_1 - A_1)\sum_{p=1}^{3}H(\eta_p)}{A_2\sum_{p=1}^{3}H(\eta_p) + B_2\left[1-\sum_{p=1}^{3}H(\eta_p)\right]}, \quad (5)$$

and

$$x_{Nd}^{\beta_1}(x_{Nd},\eta_1,\eta_2,\eta_3) = \frac{A_2 x_{Nd} + \frac{1}{2}(A_1 - B_1)\left[1-\sum_{p=1}^{3}H(\eta_p)\right]}{A_2\sum_{p=1}^{3}H(\eta_p) + B_2\left[1-\sum_{p=1}^{3}H(\eta_p)\right]}. \quad (6)$$

The energy barrier term is [18]

$$f_{barrier}(\eta_1,\eta_2,\eta_3) = W[(\eta_1^2+\eta_2^2+\eta_3^2) - 2(\eta_1^3+\eta_2^3+\eta_3^3) + (\eta_1^4+\eta_2^4+\eta_3^4) + 5(\eta_1^2\eta_2^2+\eta_2^2\eta_3^2+\eta_1^2\eta_3^2) + 5\eta_1^2\eta_2^2\eta_3^2], \quad (7)$$

where $W$ is the constant that controls the barrier heights between correspondence variants.

The gradient energy density term, which depends on the gradients of the order parameters, takes the form [18]

$$f_{grad}(\nabla\eta_1, \nabla\eta_2, \nabla\eta_3) = \frac{1}{2}\sum_{p=1}^{3} \kappa_{ij}^{p} \left(\frac{\partial \eta_p}{\partial r_i}\right)\left(\frac{\partial \eta_p}{\partial r_j}\right), \tag{8}$$

where the index, $p$, represents each variant, and $\kappa_{ij}^{p}$ is the gradient energy coefficient tensor element for the order parameter, $\eta_p$, and the variables $r_i$ and $r_j$ represent the Cartesian coordinates with $i$ and $j$ as the spatial indices. The Einstein summation notation is used for the spatial indices, in which the sum over repeated indices is implied.

Based on Eshelby's theory of inclusions [31], the elastic strain energy term expressed in terms of the order parameters proposed by Khachaturyan [32] is given by

$$f_{el}(\eta_1, \eta_2, \eta_3, \boldsymbol{u}) = \frac{1}{2} C_{ijkl}(\eta_1, \eta_2, \eta_3)\left[\epsilon_{ij}(\boldsymbol{u}) - \epsilon_{ij}^{0}(\eta_1, \eta_2, \eta_3)\right]\left[\epsilon_{kl}(\boldsymbol{u}) - \epsilon_{kl}^{0}(\eta_1, \eta_2, \eta_3)\right], \tag{9}$$

where $C_{ijkl}$ denotes an element of the stiffness tensor and $\epsilon_{ij}$ is an element of the strain tensor, which depends on the displacement vector, $\boldsymbol{u}$. The terms $\epsilon_{ij}^{0}$ are elements in SFTS. We ignore the concentration dependence of elastic properties within each phase since the precipitate is an intermetallic and the Nd concentration in the matrix is very low (≤ 0.4%). The stiffness tensor and SFTS are interpolated between the matrix and precipitate values via [18]

$$\boldsymbol{C}(\eta_1, \eta_2, \eta_3) = \sum_{p=1}^{3} \boldsymbol{C}^{\boldsymbol{\eta_p}} H(\eta_p) + \boldsymbol{C}^{\boldsymbol{\alpha}}\left[1 - \sum_{p=1}^{3} H(\eta_p)\right], \tag{10}$$

$$\boldsymbol{\epsilon}^{\boldsymbol{0}}(\eta_1, \eta_2, \eta_3) = \sum_{p=1}^{3} \boldsymbol{\epsilon}^{\boldsymbol{\eta_p}} H(\eta_p) + \boldsymbol{\epsilon}^{\boldsymbol{\alpha}}\left[1 - \sum_{p=1}^{3} H(\eta_p)\right], \tag{11}$$

where $\boldsymbol{C}^{\boldsymbol{\eta_p}}$ is the stiffness tensor for the $p^{\text{th}}$ variant, $\boldsymbol{C}^{\boldsymbol{\alpha}}$ is the stiffness tensor for the α-Mg matrix, $\boldsymbol{\epsilon}^{\boldsymbol{\eta_p}}$ is the SFTS tensor for the $p^{\text{th}}$ variant, and $\boldsymbol{\epsilon}^{\boldsymbol{\alpha}}$ is the SFTS tensor for the α-Mg matrix. The elastic strain tensor, defined in terms of the displacement vector, $\boldsymbol{u}$, is written as [18]

$$\epsilon_{ij}(\boldsymbol{u}) = \frac{1}{2}\left(\frac{\partial u_j}{\partial r_i} + \frac{\partial u_i}{\partial r_j}\right). \tag{12}$$

The mechanical equilibrium condition,

$$\frac{\partial}{\partial r_j}\left(C_{ijkl}(\eta_1,\eta_2,\eta_3)[\epsilon_{kl}(\boldsymbol{u})-\epsilon^0_{kl}(\eta_1,\eta_2,\eta_3)]\right)=0, \tag{13}$$

is solved to determine $\boldsymbol{u}$, which is used in the evolution equation, given below.

Using the variational approaches described in Refs. [33, 34], the governing equations of the composition and structural order parameter fields can be obtained [16, 18, 30]:

$$\begin{aligned}\frac{\partial x_{Nd}}{\partial t} = D\nabla\cdot\Bigg\{&\nabla x_{Nd}+\left(x^{\alpha}_{Nd}-x^{\beta_1}_{Nd}\right)\sum_{p=1}^{3}\left[\frac{\partial H(\eta_p)}{\partial \eta_p}\nabla\eta_p\right]\\ &-\left(\frac{\partial^2 f_{bulk}}{\partial x^2_{Nd}}\right)^{-1}\nabla\left[C_{ijkl}\left(\frac{\partial \epsilon^0_{ij}}{\partial x_{Nd}}\right)(\epsilon_{kl}-\epsilon^0_{kl})\right]\Bigg\},\end{aligned} \tag{14}$$

$$\begin{aligned}\frac{\partial \eta_p}{\partial t} = -L\Bigg\{&\left[f_{\beta_1}-f_\alpha-\left(x^{\alpha}_{Nd}-x^{\beta_1}_{Nd}\right)\frac{\partial f_{\beta_1}}{\partial x^{\beta_1}_{Nd}}\right]\frac{\partial H(\eta_p)}{\partial \eta_p}+\frac{\partial f_{barrier}}{\partial \eta_p}\\ &-\kappa^p_{ij}\frac{\partial^2 \eta_p}{\partial r_i\partial r_j}-C_{ijkl}\left(\frac{\partial \epsilon^0_{ij}}{\partial \eta_p}\right)(\epsilon_{kl}-\epsilon^0_{kl})\\ &+\frac{1}{2}\frac{\partial C_{ijkl}}{\partial \eta_p}\left(\epsilon_{ij}-\epsilon^0_{ij}\right)(\epsilon_{kl}-\epsilon^0_{kl})\Bigg\},\end{aligned} \tag{15}$$

where $D$ is the diffusion coefficient and $L$ is the kinetic interface mobility; both are assumed constant and isotropic in this model for simplicity.

The phase-field model in this study was parameterized by selecting thermodynamic, elastic, and kinetic parameters to capture the evolution of $\beta_1$ precipitates in the Mg–Nd alloy at 523 K, as will be discussed in section 4.3. All parameters were converted into nondimensional units for simulations using characteristic scales for energy, length, and time. Further details will be provided in section 3.3.

### 3.2. Nucleus seeding method

To simulate concurrent precipitate nucleation and growth, a stochastic method to introduce nuclei was coupled with the phase-field model. We employ the method proposed by Jokisaari et al. [35] that extended the approach by Simmons et al. [36]. According to the classical nucleation theory, the nucleation rate, $J(\boldsymbol{r},t)$, defined as the number of new nuclei formed per unit volume per unit time at the spatial position $\boldsymbol{r}$ and time $t$, is given by [37]

$$J(\boldsymbol{r},t) = ZN_n\beta \exp\left(-\frac{\Delta G^*}{kT}\right)\exp\left(-\frac{\tau_{in}}{t}\right), \tag{16}$$

where $Z$ is the Zeldovich factor, $N_n$ is the number of nucleation sites per unit volume, $\beta$ is the frequency of molecular attachment to the nucleus, $\Delta G^*$ is the nucleation energy barrier, $k$ is the Boltzmann constant, $T$ is the temperature, and $\tau_{in}$ is the incubation time. For a spherical nucleus in 3D, the nucleation energy barrier is expressed as [37]

$$\Delta G^* = \frac{16\pi\gamma^3}{3(|\Delta G_v| - |\Delta G_{strain}|)^2}, \tag{17}$$

where $\gamma$ is the interfacial energy between the matrix and precipitate phases, $\Delta G_v$ is the bulk free energy change during nucleation per unit volume of the nucleus, and $\Delta G_{strain}$ is the misfit strain energy per unit volume of the nucleus. As shown in section S3 of the SI, $\Delta G_v$ is approximately proportional to the supersaturation, $\Delta x(\boldsymbol{r},t)$, at position, $\boldsymbol{r}$, and time, $t$, defined as

$$\Delta x(\boldsymbol{r},t) = x(\boldsymbol{r},t) - x_e^\alpha, \tag{18}$$

where $x(\boldsymbol{r},t)$ is the local matrix composition and $x_e^\alpha$ is the equilibrium matrix composition. When ignoring $\Delta G_{strain}$ and taking $\Delta G_v$ as approximately proportional to $\Delta x(\boldsymbol{r},t)$, the nucleation rate can be written as [36]

$$J(\boldsymbol{r},t) = \rho_1 \exp\left(-\frac{\rho_2}{[\Delta x(\boldsymbol{r},t)]^2}\right)\exp\left(-\frac{\tau_{in}}{t}\right), \tag{19}$$

where $\rho_1$ and $\rho_2$ are combinations of physical constants (further discussed in section 4.4), and the dimensionality in equation (18) is assumed to be 3. We employ the form $\exp(-\rho_2/\Delta x^2)$ in our seeding method, even though the simulations are carried out in 2D, to preserve the quadratic dependence of the 3D nucleation energy barrier on supersaturation. By doing so, we aim to capture the nucleation behavior in a three-dimensional system in the simulation.

To examine the effect of $\Delta G_{strain}$, we numerically calculated a posteriori the ratio $|\Delta G_{strain}|/|\Delta G_v|$ of an isolated seed. We found this ratio to be less than 0.10 initially, when the supersaturation is maximum, but it becomes as large as 0.37 at the end stage of nucleation when supersaturation decreases. Nevertheless, we neglect $\Delta G_{strain}$ in equation (17) since self-consistently computing it is challenging due to the long-range nature of elastic interactions. However, as discussed in section S5 in the SI, the contribution of $\Delta G_{strain}$ to the nucleation rate

is implicitly accounted for by treating the nucleation rate parameters as effective values, which were obtained by fitting to precipitate number density data, as will be presented in section 4.4.

With the assumption that $J$ is uniform within a small volume $\Delta V$ and constant during a small time interval $\Delta t_{nuc}$, the local probability, $P(\boldsymbol{r}, t)$, that at least one nucleation event can occur in $\Delta V$ within $\Delta t_{nuc}$ can be expressed as [35, 36]

$$P(\boldsymbol{r}, t) = 1 - \exp(-J\Delta V \Delta t_{nuc}). \tag{20}$$

The nucleation probability, which varies spatially with local supersaturation and changes over time, is evaluated at fixed time intervals for every element of the mesh in the computational domain. A nucleus is introduced within the element whenever this probability exceeds a random number drawn from a uniform distribution in the interval $[0,1)$. Since our experiments show no evidence of preferential orientations (see section 4.1), we randomly selected the correspondence variant for nucleation. The order parameter for that variant was then modified within a small elliptical region centered on the nucleation site, with its major axis along one of the two primary possible growth directions, which is also chosen at random. We found that elongating the seed along the selected growth direction is sufficient to direct the growth of the plate. A short holding period is imposed after seeding the nucleus [35]. During the holding period, the structural order parameter near the nucleation site is maintained, thereby preserving the driving force for initial solute accumulation and preventing the immediate dissolution of the newly formed nucleus.

While this nucleus seeding algorithm provides a stochastic method for introducing nuclei that takes into account the local conditions, many of its parameters are not directly obtainable or have significant uncertainties such as the Zeldovich factor, molecular attachment frequency, geometric factor (in the case of heterogeneous nucleation), interfacial energy (which appears with third power), equilibrium compositions in each phase, and supersaturation as a function of time and space. Thus, we employ a multistep parameterization procedure summarized below and described in detail in section 4.4. Experimental TEM images were processed to estimate 3D number densities, which were then converted to 2D for use in 2D simulations, by maintaining the same characteristic interparticle distances. Initial optimization of nucleation rate parameters was performed to minimize discrepancies between the experimental number densities and the model that assumes uniform supersaturation. The simulation was then conducted with this set of parameters. To correct for the limitations inherent in the uniform-supersaturation model, the simulation time scale is adjusted, which is equivalent to scaling $\rho_1$, $D$, and $\tau_{in}^{-1}$ with the same factor, by optimizing the match between the corrected simulation results and experimental data. This workflow ensures that the modeling framework is robustly calibrated to physical realities while maintaining computational efficiency. It is important to note that we ignore the intermediate phases that are involved in the formation of $\beta_1$ precipitates. This approximation allows us to employ a simplified model that considers a single effective nucleation barrier and to avoid tracking the populations of intermediate precipitate phases, which would introduce additional model parameters and increase computational cost. Thus, the nucleation parameters determined in this work are effective, empirical values that represent the rate-limiting step.

### 3.3. Numerical implementation

The governing equations (equations (13)–(15)) were nondimensionalized and implemented in PRISMS-PF [38], an open-source high-performance phase-field framework, using an unstructured mesh with quadrilateral second-order elements. The characteristic energy density, $f^* = 2 \times 10^9$ J m$^{-3}$ $= 2 \times 10^{-18}$ J nm$^{-3}$, and the characteristic length, $l^* = 10^{-9}$ m $= 1$ nm, were used for nondimensionalization. The dimensionless diffusivity was chosen to be 1, which sets the time scale to be $t^* = l^{*2}/D$; $D$ is determined in the parameterization process described in section 4.4. All simulations were performed in a square domain with a side length of $1100l^*$. The side length of square-shaped elements ranges from $1100l^*/2^{12}$ to $1100l^*/2^6$, which is adjusted dynamically during the simulation based on the adaptive-mesh-refinement criteria. The interfacial width was resolved by approximately 5.6 second-order elements. The time step was chosen to be $2.5 \times 10^{-3}t^*$, which yielded numerical stability. A nucleus seeding attempt was made every $2.5t^*$, which gives $\Delta t_{nuc}$ in equation (20); this value is much smaller than the characteristic time scale for nucleation at the peak rate (on the order of $10^3t^*$). The seed has semi-major and semi-minor axis lengths of $3l^*$ and $2l^*$, respectively. The holding period after seeding a nucleus was set to $100t^*$, which was found to be sufficient to stabilize a seed. A solid solution of uniform Nd composition of 0.004 in mole fraction was set as the initial condition, and periodic boundary conditions were employed. A summary of model parameters is presented in table S1 of the SI.

## 4. Results and discussions

### 4.1. Determination of precipitate number density

This study focuses specifically on $\beta_1$ precipitates in Mg–Nd alloys within the precipitation sequence [8]: SSSS → GP zones → $\beta'''$ precipitates (a hybrid ordering of $\beta'$ with a $Mg_7Nd$ composition and $\beta''$ with a $Mg_3Nd$ composition) → $\beta_1$ precipitates ($Mg_3Nd$) → $\beta$ precipitates ($Mg_{12}Nd$) → $\beta_e$ precipitates ($Mg_{41}Nd$). The last two phases were not observed under the experimental condition we investigated. The $\beta_1$ precipitates exhibit a plate-like morphology, with a habit plane of $\{1\bar{1}00\}_\alpha$ in the matrix. The orientation relationship between the $\beta_1$ precipitates and the α-Mg matrix is $\{\bar{1}12\}_{\beta_1}//\{1\bar{1}00\}_\alpha$ and $\langle 110\rangle_{\beta_1}//[0001]_\alpha$ [15]. The morphologies and spatial distributions of $\beta_1$ precipitates at various aging times are shown in figure 2. The TEM foil samples were oriented with the normal direction along [0001] of the α-Mg matrix. Theoretically, under the bright-field STEM mode, all six correspondence and rotation variants are expected to yield comparable TEM intensities, and they are expected to be captured equally likely in the TEM images. Indeed, the number fractions of the three growth directions are approximately 1/3 each. The deviation of the three fractions from 1/3 calculated for each TEM image varies from 0.01 to 0.20. This scatter may arise from the limited sampling area and the statistics associated with a small number of precipitates in the TEM images, as well as from non-uniformity in local TEM sample thickness, which could cause contrast variation [39]. In fact, the deviation was especially pronounced in images with poor image quality (e.g., figure 2(c)). Thus, we believe that no single primary growth direction was strongly favored during aging.

The estimated 3D number density for each sample was calculated from the measured length along the precipitate's major axis in a TEM image, $d_{m,i}$, where $i$ indexes the precipitates. The measurements were made using ImageJ [40] by manually overlaying each precipitate with an ellipse with similar dimensions. The estimated 3D number density from each TEM image was obtained by using [29]

$$N_V^{exp} = \frac{1}{A_{foil}} \sum_{i=1}^{N'} \frac{1}{t_{foil} + d_{m,i}}, \tag{21}$$

where the sum is over all precipitates intersecting the foil. The subscript $V$ indicates that the quantity is per unit volume, $t_{foil}$ is the foil thickness, and $A_{foil}$ is the sample area (see also section

S1 of the SI). Table 1 summarizes the measured sample area, foil thickness, and the corresponding estimated 3D number density for each TEM image shown in figure 2.

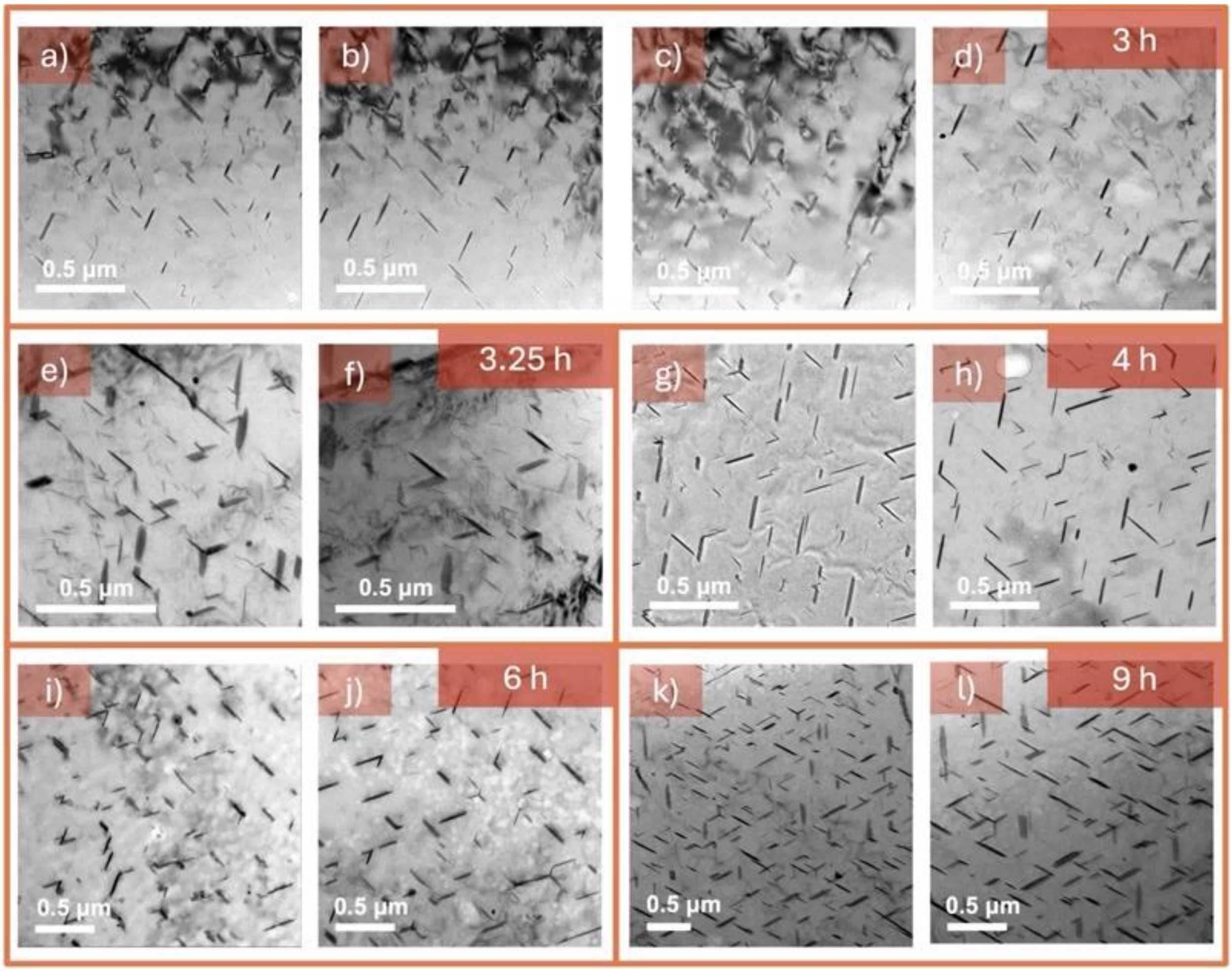

**Figure 2.** TEM images showing the microstructures of $\beta_1$ precipitates in Mg–Nd alloys at various aging times: (a–d) 3 h, (e–f) 3.25 h, (g–h) 4 h, (i–j) 6 h, and (k–l) 9 h, all at 523 K. The normal direction of all foils was oriented along [0001] of the α-Mg matrix. Note that the scale bars vary in size across the subfigures.

**Table 1.** Measured sample area, foil thickness, and estimated 3D number density for each TEM image in figure 2.

| **Figure 2 panel** | **Aging time $t$ (h)** | **Sample area $A_{foil}$ (μm$^2$)** | **Foil thickness $t_{foil}$ (μm)** | **3D number density $N_V^{exp}$ (count μm$^{-3}$)** |
|---|---|---|---|---|
| (a) | 3 | 2.56 | 0.158 | 87.9 |
| (b) | 3 | 2.56 | 0.158 | 83.8 |
| (c) | 3 | 2.56 | 0.158 | 82.0 |
| (d) | 3 | 2.56 | 0.158 | 81.2 |
| (e) | 3.25 | 1.40 | 0.224 | 125.6 |
| (f) | 3.25 | 1.40 | 0.224 | 115.4 |
| (g) | 4 | 2.56 | 0.182 | 162.6 |
| (h) | 4 | 2.56 | 0.182 | 158.5 |
| (i) | 6 | 5.82 | 0.141 | 79.6 |
| (j) | 6 | 5.82 | 0.141 | 78.0 |
| (k) | 9 | 10.19 | 0.377 | 64.1 |
| (l) | 9 | 5.83 | 0.377 | 69.3 |

## 4.2. Determination of matrix composition

To further elucidate the microstructural evolution during aging, the composition of the matrix was assessed using APT. Composition measurements of the alloying element were obtained across the matrix/precipitate interface with a bin size of 0.5 nm for three different aging times: 4 h, 6 h, and 100 h. APT datasets were acquired and analyzed from 2 to 4 APT specimens for each aging time. The matrix compositions were determined by analyzing matrix regions located sufficiently away from the interface. For dilute species detected via APT, such as Nd atoms in the matrix, the statistical uncertainty in the measured composition can be estimated using Poisson statistics [41, 42]

$$\sigma = \sqrt{\frac{x_{Nd}(1 - x_{Nd})}{C_{total}}}, \tag{22}$$

where $\sigma$ is the statistical uncertainty in Nd composition, $x_{Nd}$ is the measured Nd composition, and $C_{total}$ is the total atom count. The APT measurements and corresponding statistical uncertainties are summarized in table 2. The data at 100 h are included in the table to demonstrate that, by 6 h, the matrix composition has approached the asymptotic limit, and they are not used in later parameterization or results. The Poisson uncertainty values, which represent the lower bounds for the measurement error, are one to two orders of magnitude smaller than the corresponding Nd mole fractions, highlighting the high precision of atom counting. However, the variations observed between repeated measurements in table 2 are larger than the statistical uncertainties. This implies substantial contributions from other sources of measurement uncertainty, such as systematic factors like detector efficiency, as well as actual spatial variations [43, 44].

**Table 2**. Summary of APT measurements of matrix Nd composition (expressed as mole fraction) and corresponding statistical uncertainties at different aging times.

| Aging times | Measurement | Nd atoms / total atoms detected | Mole fraction of Nd ($x_{Nd}$) | Statistical uncertainty ($\sigma$) |
|---|---|---|---|---|
| 4 h | 1 | 274/869046 | $3.15 \times 10^{-4}$ | $1.90 \times 10^{-5}$ |
| | 2 | 35/82308 | $4.25 \times 10^{-4}$ | $7.19 \times 10^{-5}$ |
| | 3 | 22/61318 | $3.59 \times 10^{-4}$ | $7.65 \times 10^{-5}$ |
| | 4 | 123/341201 | $3.60 \times 10^{-4}$ | $3.25 \times 10^{-5}$ |
| 6 h | 1 | 4309/11530212 | $3.74 \times 10^{-4}$ | $5.69 \times 10^{-6}$ |
| | 2 | 66/223655 | $2.95 \times 10^{-4}$ | $3.63 \times 10^{-5}$ |
| 100 h | 1 | 261/908410 | $2.87 \times 10^{-4}$ | $1.78 \times 10^{-5}$ |
| | 2 | 328/1598256 | $2.05 \times 10^{-4}$ | $1.13 \times 10^{-5}$ |
| | 3 | 1915/5190376 | $3.69 \times 10^{-4}$ | $8.43 \times 10^{-6}$ |

### 4.3. Phase-field model parameters

The chemical free energy densities of the α-Mg matrix and $\beta_1$ precipitate phases used in this work were obtained using the CALPHAD method. We surveyed several CALPHAD studies on the Mg–Nd system and found that the equilibrium compositions reported for the matrix and $\beta_1$ phases at 523 K vary across different thermodynamic databases, as summarized in table 3. The Mg–Nd thermodynamic database from Qi et al. [45] was selected for this study, as it yields simulated matrix compositions that are in closest agreement with our experimental APT measurements, specifically the asymptotic value of matrix composition at late aging times. The CALPHAD free energy curves for the α-Mg matrix and $\beta_1$ phases are presented in figure 3(a), along with the constrained quadratic fits reconstructed by including the common-tangent linear parts in the quadratic fits in figure 3(b). Thermodynamically equivalent free energy functions were obtained by subtracting the common tangent line from the original CALPHAD curves. As shown in figure 3(b), these adjusted free energies were fitted to quadratic functions around the equilibrium compositions. This fitting procedure ensures that the equilibrium concentrations are retained in the resulting fitted free energy densities. The coefficients obtained for the fitted parabolic functions in equations (3) and (4) are: $A_0 = 4.00\times10^{-1}$ J mol$^{-1}$, $A_1 = -4.23\times10^{3}$ J mol$^{-1}$, $A_2 = 1.12\times10^{7}$ J mol$^{-1}$, $B_0 = 1.71\times10^{4}$ J mol$^{-1}$, $B_1 = -1.41\times10^{5}$ J mol$^{-1}$, and $B_2 = 2.89\times10^{5}$ J mol$^{-1}$.

To confirm the validity of the quadratic fits and address the potential for auxiliary composition deviation within the nominally stoichiometric $\beta_1$ phase, the local auxiliary composition $x_{Nd}^{\beta_1}$ was calculated within the bulk and interfacial regions (sum of order parameters greater than 0.1) for the simulation presented in figure 6 and figure 7. The average relative deviation from the equilibrium value was 1.4% (corresponding to $\Delta x_{Nd,ave}^{\beta_1} \approx 0.003$), with a maximum deviation of 8.2% (corresponding to $\Delta x_{Nd,max}^{\beta_1} \approx 0.02$). The higher values of $x_{Nd}^{\beta_1}$ were observed at the tips of precipitates or within disappearing precipitates, which include a significant contribution from the Gibbs–Thomson effect. Flatter interfaces had values very close to the equilibrium value (~1.3% deviation), indicating that the quadratic fit did not allow for large deviations. On the other hand, $x_{Nd}^{\alpha}$ exhibited a substantial relative deviation with respect to its equilibrium value (~50% at maximum); however, this was due to the extremely small value of $x_e^{\alpha} = 0.00019$. Because the auxiliary compositions appear only in the governing equation as the difference between the two phases, $\left(x_{Nd}^{\alpha} - x_{Nd}^{\beta}\right)$, we evaluated this quantity within the interfacial regions (sum of order parameters between 0.1 and 0.9). The average relative deviation of the local auxiliary composition difference from that of the equilibrium values, $\left(x_e^{\alpha} - x_e^{\beta_1}\right)$, was 1.5%. The larger deviations again occurred at the tips or in precipitates that were disappearing with a maximum value of 8.0%, also most likely due to the Gibbs–Thomson effect. Overall, these results confirm that the fitting procedure does not introduce significant error.

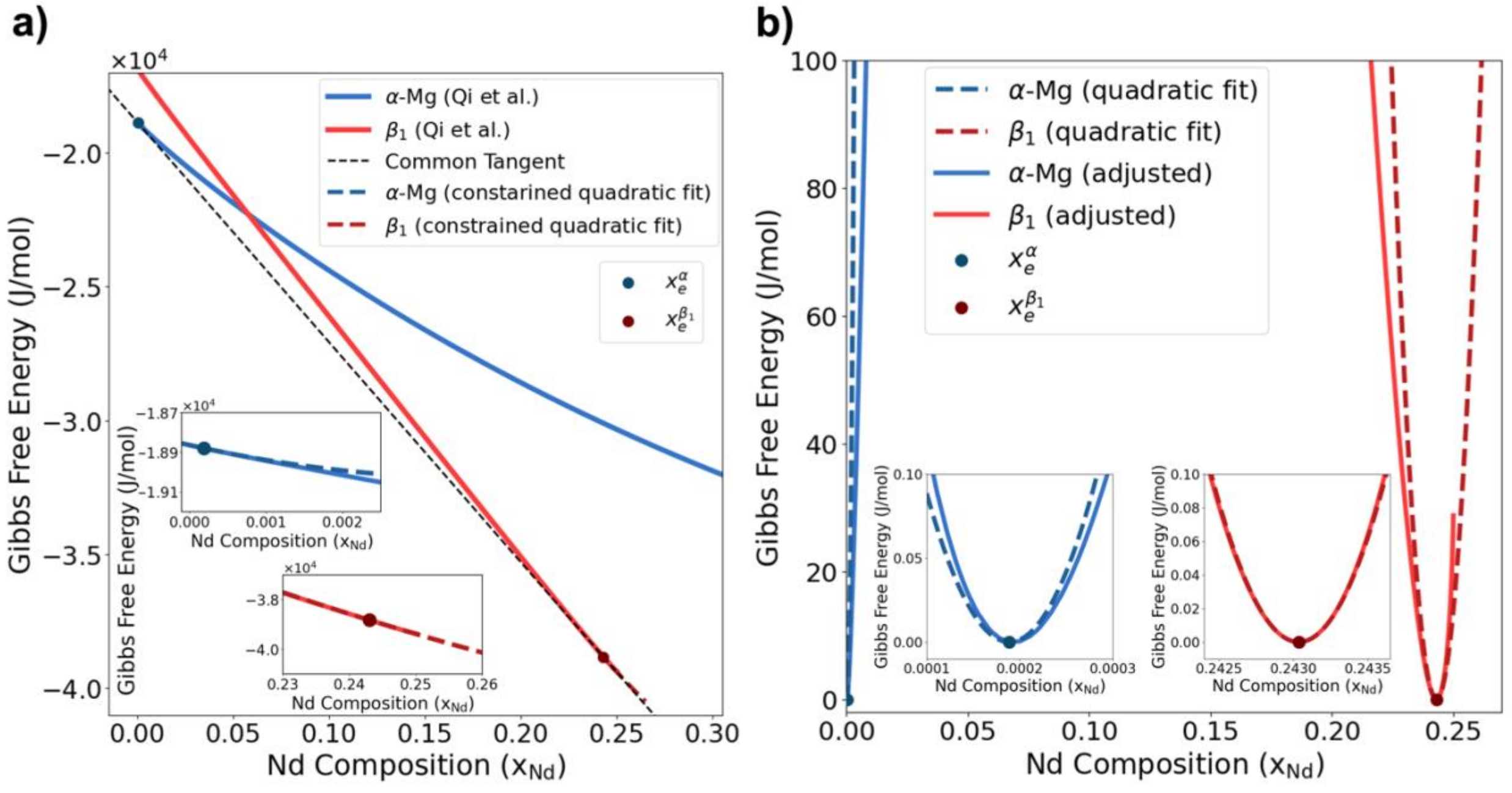

**Figure 3.** (a) Molar Gibbs free energy of the α-Mg matrix and $\beta_1$ phases plotted as a function of Nd composition at 523 K, calculated using the thermodynamic parameters from Qi et al. [45]. The dots indicate equilibrium compositions. The insets show magnified views near the equilibrium compositions, along with the constrained quadratic fits from (b) but including the common-tangent linear parts. (b) Adjusted Gibbs free energy curves for the two phases, obtained by subtracting the common tangent from the original curves, with quadratic fits applied near the equilibrium compositions (shown as dots). The insets in (a) and (b) provide magnified views near the equilibrium compositions for both phases.

The interfacial energy between the $\beta_1$ and hcp α-Mg phases is assumed to be isotropic in this study for simplicity. Thus, we take $\kappa_{ij} = K\delta_{ij}$ in equation (8), where $\delta_{ij}$ is the Kronecker delta and $K$ is the gradient energy coefficient. An interfacial energy of $\gamma = 0.05$ J m$^{-2}$ was selected to maintain consistency with a previous phase-field study [17] due to the lack of experimental values in the literature. The interfacial energy in this model is related to the phase-field model parameters by [16, 30]

$$\gamma = \sqrt{KW}/\left(3\sqrt{2}\right), \tag{23}$$

where $K = 5.115 \times 10^{-2} f^*/l^{*2}$ and $W = 0.22 f^*$ is the constant in the energy barrier term in equation (7). These constants were chosen to yield the value of $\gamma$ reported above, and an equilibrium interface width (measured from $\eta_p = 0.1$ to 0.9) of $\delta \approx 2.2\sqrt{2K/W} = 1.5 l^*$, which is resolved by approximately 5.6 second-order elements, as mentioned in section 3.3. A diffusivity value of $D = 7.5 \times 10^{-18}$ m$^2$ s$^{-1}$, which is similar to a value for Nd in pure Mg reported in Ref. [46], was adopted in the simulation before fine-tuning model parameters. The fine-tuning of the time scale required to optimize the agreement between the simulation and experiment was equivalent to using a diffusivity of $D = 10.4 \times 10^{-18}$ m$^2$ s$^{-1}$, as will be discussed in section 4.4.

**Table 3.** Summary of equilibrium Nd compositions in the matrix phase and $\beta_1$ phase at 523 K obtained from different thermodynamic databases.

| Mg–Nd system thermodynamic database | Equilibrium Nd mole fraction in the matrix at 523 K | Equilibrium Nd mole fraction in the $\beta_1$ precipitates at 523 K |
|---|---|---|
| Niu et al. 2010 [47] | $1.00 \times 10^{-14}$ | $2.35 \times 10^{-1}$ |
| Qi et al. 2011 [45] | $1.89 \times 10^{-4}$ | $2.43 \times 10^{-1}$ |
| Kang et al. 2012 [48] | $1.56 \times 10^{-4}$ | $2.14 \times 10^{-1}$ |
| Guo et al. 2013 [49] | $7.20 \times 10^{-5}$ | $2.35 \times 10^{-1}$ |
| Luo et al. 2025 [50] | $3.56 \times 10^{-4}$ | $2.23 \times 10^{-1}$ |

The elastic constants of the $\beta_1$ precipitates are assumed to be identical to those of the α-Mg matrix phase, as in Ref. [17]. The following reported elastic constants for the matrix phase were used [16]: $C_{1111} = C_{2222} = 62.6$ GPa, $C_{3333} = 64.9$ GPa; $C_{1122} = 26.0$ GPa; $C_{1133} = C_{2233} = 20.9$ GPa; $C_{2323} = C_{1313} = 13.3$ GPa; and $C_{1212} = 18.3$ GPa, where the remaining elements are obtained from the symmetry relations, $C_{ijkl} = C_{jikl} = C_{ijlk}$, or equal to zero. The SFTS tensors for each variant are given by

$$\epsilon^{\eta_1} = \begin{bmatrix} 0.1111 & 0.0962 & 0 \\ 0.0962 & 0 & 0 \\ 0 & 0 & 0 \end{bmatrix}, \quad (24)$$

$$\epsilon^{\eta_2} = \begin{bmatrix} -0.0556 & 0 & 0 \\ 0 & 0.1666 & 0 \\ 0 & 0 & 0 \end{bmatrix}, \quad (25)$$

$$\epsilon^{\eta_3} = \begin{bmatrix} 0.1111 & -0.0962 & 0 \\ -0.0962 & 0 & 0 \\ 0 & 0 & 0 \end{bmatrix}, \quad (26)$$

where the SFTS in the out-of-plane direction is ignored.

### 4.4. Parameterization workflow

The model parameterization workflow involved the following steps, which are further detailed:

1) Obtain the effective 2D number density from the estimated 3D number density presented in section 4.1 to 2D for phase-field simulations.
2) Perform an initial optimization of parameters $\rho_1$, $\rho_2$ and $\tau_{in}$ in equation (19), which determine the nucleation rate.
3) Find a diffusivity value that yields reasonable agreement between simulated and effective 2D number density evolution. Conduct the simulation using this diffusivity and the nucleation rate parameters found in step 2.
4) Fine-tune the simulation time scale by further optimizing the agreement between simulated and the effective 2D number density evolution.

*Step 1. Obtain the effective 2D number density from the 3D number density*

For computational efficiency, the phase-field simulations were performed in 2D rather than 3D. Each reported 2D simulation required up to $1.5 \times 10^7$ degrees of freedom, which is reached near the end of the nucleation stage, and used approximately 30 wall-clock hours using 1024 CPU cores. The high computational cost stems from the need to resolve the diffuse interfaces that represent highly anisotropic precipitate shapes and large curvatures. A comparable 3D simulation at the same spatial resolution would increase the degrees of freedom (and the computational cost) by at least two orders of magnitude, making the 3D simulation impractical for the present study. However, a key discrepancy exists between 2D and 3D spatial dimensions regarding particle interactions. To reconcile this, we applied a scheme to obtain the effective 2D number density from the estimated 3D number density data that preserves the characteristic interparticle distance over which diffusion must occur during coarsening. To establish the relationship, we assume a statistically uniform spatial distribution of precipitates and the same characteristic interparticle distance, $\lambda$, both in 2D and 3D. Therefore, the number density per unit volume in 3D, $N_V = \lambda^{-3}$, and the number density per unit area in 2D, $N_A = \lambda^{-2}$, can be related by

$$N_A = {N_V}^{2/3}, \tag{27}$$

where the subscript $A$ indicates that the quantity is per unit area. We employ this relationship to obtain the effective 2D number densities for different aging times, $N_A^{eff}(t)$, to parameterize the phase-field model, as described in the following steps.

*Step 2. Perform initial optimization of nucleation rate parameters*

To quantitatively model nucleation behavior in this system, it is necessary to set the three parameters, $\rho_1$, $\rho_2$ and $\tau_{in}$, that determine the nucleation rate via equation (19). As mentioned earlier, these parameters cannot be uniquely determined directly from the available experimental data. However, an initial optimization can be performed by (1) selecting initial guesses for the values of $\rho_1$, $\rho_2$ and $\tau_{in}$ within a set of bounds based on either models or experimental observations, (2) integrating the nucleation rate over time during the nucleation stage to calculate the number density, $N_A(t)$, and (3) optimizing $\rho_1$, $\rho_2$ and $\tau_{in}$ to minimize the difference between $N_A(t)$ and $N_A^{eff}(t)$.

For this initial optimization, we ignore the spatial variation in supersaturation and treat it as a function of time only. The limitations of this uniform-supersaturation model are addressed using the fine-tuning method discussed in step 4. The time dependence of supersaturation is accounted for by modeling the Nd composition in the matrix, $x$, as a function of aging time using a hyperbolic tangent function. This function captures how the supersaturation decreases rapidly during the nucleation stage and then transitions to the equilibrium value as it enters the coarsening stage. The APT measurements are used to set the initial and asymptotic values of the matrix composition (table 2 and figure S3(a)), while the transition time, $t_{trans}$, and width, $\tau_w$, are estimated from the number density data measured from the micrograph (table 2 and figure S3(a)). The resulting function,

$$x_{fit}(t) = x_e + (x_0 - x_e)\left[\frac{1}{2} - \frac{1}{2}\tanh\left(\frac{t - t_{trans}}{\tau_w}\right)\right], \quad (28)$$

is shown in figure S3(a) in the SI. Here, $x_e = 0.0003$ is the asymptotic value of $x_{fit}(t)$ at late aging times, $x_0 = 0.004$ is the initial Nd composition in the solid, $t_{trans} = 2.75$, and $\tau_w = 0.88$. From this function, the supersaturation, $\Delta x_{fit}(t) = x_{fit}(t) - x_e$, was obtained. Then, the number density as a function of time can be estimated by

$$N_A(t) = \int_0^t J\big(\Delta x_{fit}(t'), t'; \rho_1, \rho_2, \tau_{in}\big)dt'. \quad (29)$$

The search bounds for each nucleation rate parameter (listed in table 4) were determined as follows. In classical nucleation theory, the incubation time, $\tau_{in}$, represents the characteristic time before the nucleation rate reaches steady state. As shown in figure 4(a), the experimental number density increases significantly between 2.0 h and 3.0 h. Therefore, a search range from 0.0 to 3.0 h was established for finding the optimal value of $\tau_{in}$. The optimization yielded an incubation time at the upper bound of the allowed interval (approximately 3.0 h). When a broader interval (0.0–10.0 h) was chosen for the optimization, the best-fit value shifted to approximately 6.9 h. However, this value provided only a minor improvement to the fit and is inconsistent with the assumption that the time to reach steady-state nucleation is determined by $\tau_{in}$. Therefore, $\tau_{in}$ = 3.0 h was used as the incubation time.

For $\rho_2$, we used an approximate value given by

$$\rho_2 = \frac{4\pi\gamma^3 {V_m}^2}{3kT(A_2)^2\left(x_e^{\beta_1} - x_e^{\alpha}\right)^2}, \quad (30)$$

where $V_m$ is the molar volume, and $x_e^{\alpha}$ and $x_e^{\beta_1}$ are the equilibrium Nd compositions in the matrix and $\beta_1$ phases, respectively. The derivation for this expression follows a similar approach to that of Ref. [51], and is presented in section S4 of the SI. The value of $\rho_2$ depends on several physical parameters, each with an associated uncertainty. For $x_e^{\alpha}$ and $x_e^{\beta_1}$, ranges of $1 \times 10^{-5}$–$5 \times 10^{-4}$ and 0.2–0.25, respectively, were chosen based on the survey summarized in table 3. For $\gamma$, only two values were found in the literature: 0.098 J m$^{-2}$ reported from DFT calculations [52], and 0.05 J m$^{-2}$ used in Ref. [17]. Considering the two $\gamma$ values and the ranges for $x_e^{\beta_1}$ and for $x_e^{\alpha}$, the calculated $\rho_2$ falls within a range between $2.84 \times 10^{-6}$ and $1.45 \times 10^{-5}$. Given that the range of $\rho_2$ corresponding to these $\gamma$ values spans more than one order of magnitude, this demonstrates pronounced sensitivity of $\rho_2$ to the value of $\gamma$. To ensure that the optimization process can adequately explore this uncertainty, the lower and upper bounds of $\rho_2$ were expanded to $1.0 \times 10^{-7}$ and $1.0 \times 10^{-4}$, respectively. This broader optimization range minimizes the risk of artificially constraining the parameter space and thereby enables more robust identification of optimal values.

The optimization bounds for $\rho_1$ were established starting from an order-of-magnitude estimate of the steady-state nucleation rate, $J_{ss}$, assuming it was achieved in the experiment, and expanding the range from that value. According to equation (19), in the limit $t \gg \tau$, $J_{ss} \approx \rho_1 \exp[-\rho_2(\Delta x)^{-2}]$. If the supersaturation, $\Delta x$, is sufficiently large (such that $\Delta x^2 \gg \rho_2$), the exponential term $\exp[-\rho_2(\Delta x)^{-2}]$ approaches unity, so $J_{ss} \approx \rho_1$. As shown by the experimental data in figure 4(a), the number density increases by approximately $10^{13}$ m$^{-2}$ between 2.0 h and 3.0 h of aging, giving $J_{ss} \approx 10^{13}$ m$^{-2}$ h$^{-1}$ on the order of magnitude (see figure S3(b)). Given the uncertainty in this approximation, the optimization bounds for $\rho_1$ were chosen to span two orders of magnitude above and below $10^{13}$ m$^{-2}$ h$^{-1}$.

Global optimization was performed within these bounds using the differential evolution algorithm [53] from the Python package SciPy [54]. The effective 2D number density starts to decrease after 4.0 h, which is taken as the onset of the coarsening stage. Therefore, only data points before this time were used for nucleation rate parameter optimization. The resulting optimized nucleation rate parameters of this initial step are listed in table 4. The evolution of number density calculated from the classical nucleation theory based on these parameters is plotted as the orange curve in figure 4(a).

**Table 4.** Nucleation rate parameter bounds used in parameter optimization and resulting optimized values.

| Nucleation rate parameter | $\rho_1$ ($m^{-2}$ $h^{-1}$) | $\rho_2$ | $\tau_{in}$ (h) |
|---|---|---|---|
| **Optimization bound** | $[10^{11}, 10^{15}]$ | $[1.0 \times 10^{-7}, 1.0 \times 10^{-4}]$ | $[0.0, 3.0]$ |
| **Optimized value** | $4.61 \times 10^{13}$ | $1.74 \times 10^{-7}$ | 3.0 |

*Step 3. Obtain an estimated value of diffusivity*

In the absence of precise experimental or theoretical data for Nd diffusivity in the hcp α-Mg matrix phase at 523 K, we conducted exploratory phase-field simulations based on this diffusivity value. We started with a reported diffusivity coefficient for Nd atoms in the pure hcp-Mg phase of $7.0 \times 10^{-18}$ $m^2$ $s^{-1}$ at 523 K. This value was calculated via a temperature-dependent Arrhenius equation reported in Ref. [46] and obtained experimentally from Electron Probe Microanalysis–Wavelength-Dispersive X-ray Spectroscopy (EPMA-WDS) measurements at different temperatures. Using the initially optimized nucleation rate parameters, we tested three additional diffusivity values: $6.5 \times 10^{-18}$ $m^2$ $s^{-1}$, $7.5 \times 10^{-18}$ $m^2$ $s^{-1}$, and $8.0 \times 10^{-18}$ $m^2$ $s^{-1}$. Out of these simulations, that with $7.5 \times 10^{-18}$ $m^2$ $s^{-1}$ yielded the best agreement with the effective 2D number densities converted from the 3D experimental values, specifically the maximum number density and the rate of decrease in the number density during the coarsening stage. Thus, the result from this simulation was used for the fine-tuning step below. Despite our exploration in diffusivity values, the predicted number density shows a noticeable discrepancy from the simulated results (green curve in figure 4(a)). As discussed below, this discrepancy can be further reduced by fine-tuning the model parameters.

*Step 4. Fine-tune the model parameters*

Recognizing the limitations in the initial optimization of nucleation parameters based on the uniform-supersaturation model, a method is proposed to correct model parameters. By observing that a change in the time scale significantly improves agreement between the experimental data and the simulation prediction, we determine a correction factor ($\xi$) to the time scale. This is equivalent to adjusting the model parameters such that $\rho_1' = \rho_1 \cdot \xi$, $\tau_{in}' = \tau_{in}/\xi$ and $D' = D \cdot \xi$. During the nucleation stage, both the parameterized nucleation rate and the diffusivity affect the rate of change in number density. The optimal value of $\xi$ was determined by minimizing an error metric defined as the integrated absolute difference between the trial number density curves (calculated using various $\xi$ values) and the predicted number density curve during the nucleation and growth stages (defined as 0 to 4 hours). As shown in figure 4(b), a third-order polynomial was fitted to the plot of error vs. $\xi$ to identify the minimum error, which was found at $\xi$ = 1.39, as shown in figure 4(b). To ensure statistical reliability, two additional independent simulations, besides the one shown in figure 4, were conducted using the same initially optimized nucleation rate parameters and analyzed in the same manner to obtain their optimal $\xi$ values. The three independent simulations yielded optimal $\xi$ values of 1.39, 1.46, and 1.39, with a mean of 1.41 and a low standard deviation of 0.033, demonstrating good statistical consistency.

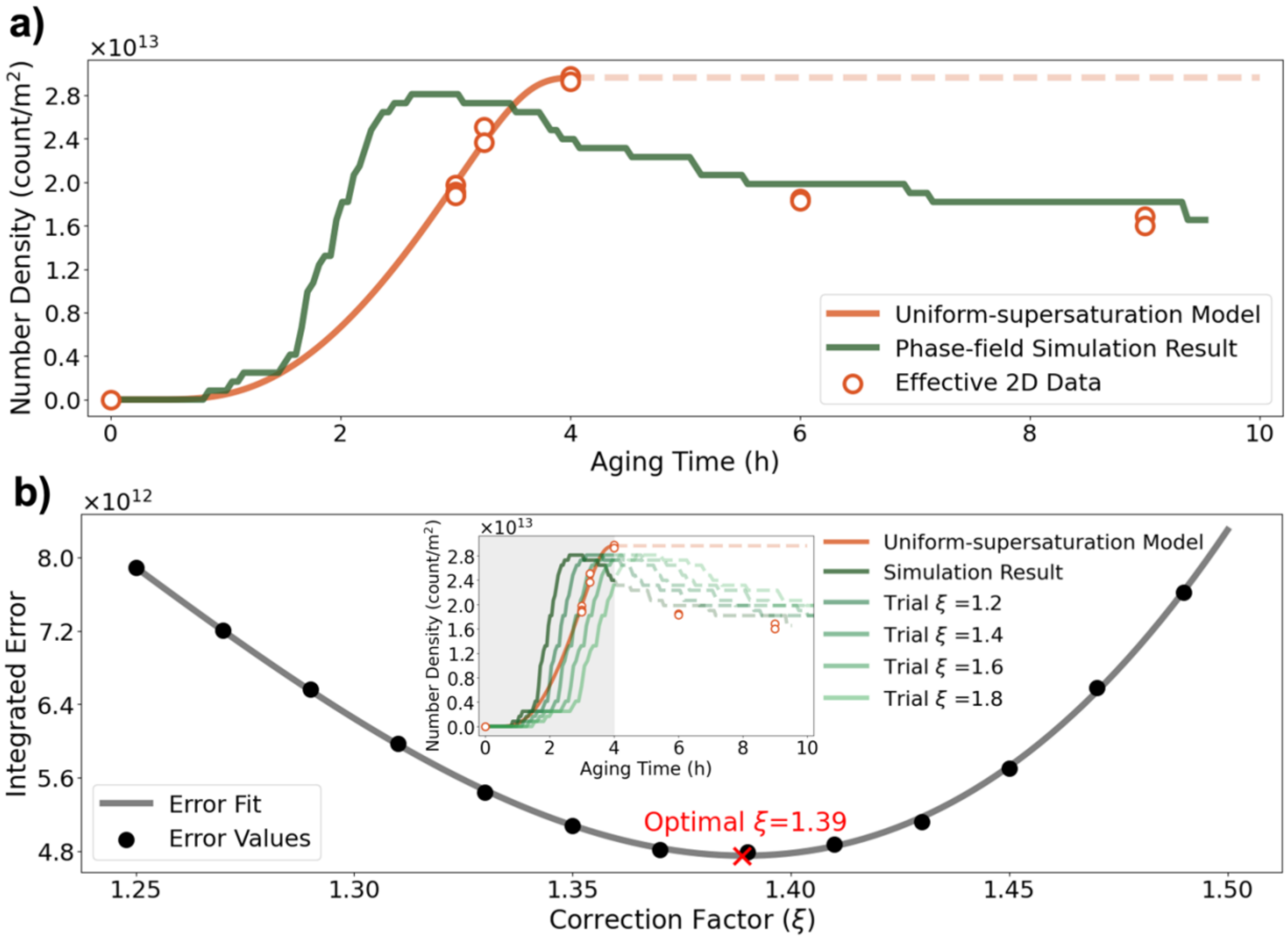

**Figure 4.** (a) Predicted number density versus aging time by the phase-field model before correction (green curve), in comparison with experimental data (open orange circles) and the uniform-supersaturation model's nucleation curve (orange curve) optimized against the experimental results. (b) Absolute differences between various trial-corrected number densities and the uniform-supersaturation model prediction that closely agree with experiment, integrated up to 4 hours (gray shaded in the inset), plotted as a function of correction factor ($\xi$). The gray curve is a third-order polynomial fit used to determine the optimal $\xi$, which corresponds to the minimum integration error. The subplot displays the number density as a function of time as in (a), with trial-corrected number density curves shown in various shades of green.

### 4.5. Simulation results and comparison with experimental data

After fine-tuning the model parameters, the simulations produce a quantitative prediction of precipitate number density as a function of aging time, in reasonable agreement with the effective 2D number densities calculated from the experimental data, as shown in figure 5. Three independent simulations were conducted using the same initially optimized nucleation rate parameters but using different sets of random numbers, each corrected by its optimal $\xi$. The results are statistically consistent, as shown by the simulation error ranges (green shades) in figure 5(a)–(c). Several representative aging times are labeled as $t_0$–$t_8$ in figure 5, which correspond to 0.00 h, 1.47 h, 2.03 h, 2.52 h, 3.01 h, 3.22 h, 3.99 h, 6.03 h, and 9.04 h, respectively. These simulation times represent the closest match to the times at which TEM measurements were obtained.

In figure 5(a), it can be observed that the precipitate number density increases significantly between $t_1$ to $t_5$ and decreases after $t_6$, similarly to the effective 2D number densities (open orange circles). The increase in number density is attributed to nucleation at early times, followed by a decrease due to coarsening. As shown in figure 5(a), some discrepancies with respect to the experimental data remain; for example, the maximum precipitate number density in the simulation falls below the maximum observed in experiments. These discrepancies likely arise from the fact that the model is in 2D and is parameterized with effective 2D number densities derived from estimated 3D experimental data. As a result, 3D diffusional interactions between precipitates may not be fully captured. In addition, using the uniform-supersaturation model to conduct the initial optimization of the nucleation parameters and then applying a single time-scale correction factor to the kinetic parameters of the model, $D$, $\rho_1$, and $\tau_{in}$, overconstrain the parameter space for optimization, resulting in a suboptimal parameter set.

Figure 5(b) shows the simulated precipitate area fraction as a function of time. The total area of precipitates was calculated by numerically integrating the sum of the order parameters over the simulation domain. As expected, when coarsening becomes dominant after $t_6$, the precipitate area fraction remains nearly constant. A comparison of the average matrix Nd composition with the APT measurement is presented in figure 5(c). For the purpose of calculating the average matrix composition, only regions where the sum of order parameters is less than 0.001 are considered. The simulated Nd compositions fall within the statistical uncertainties of all but one of the experimental measurements.

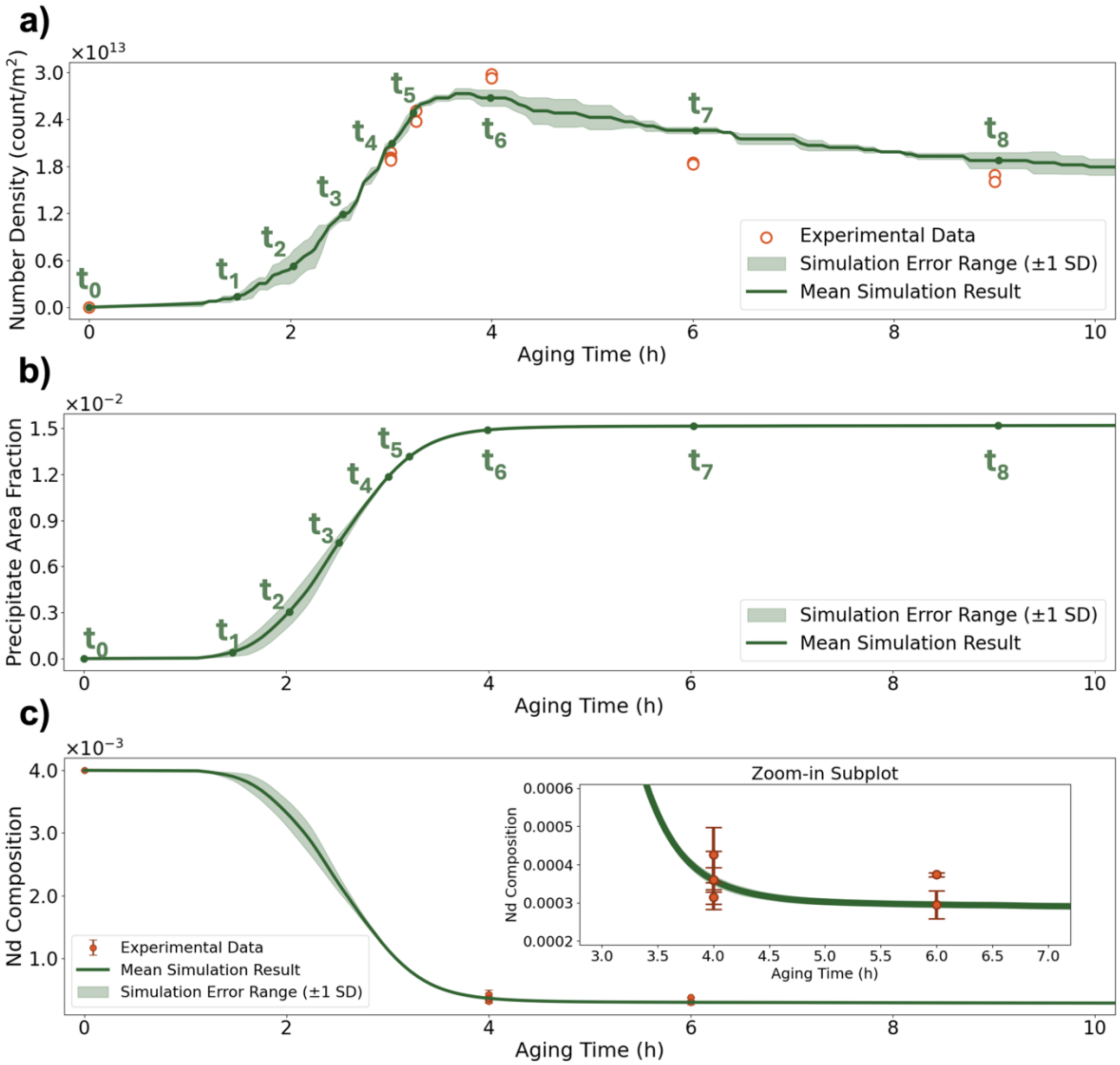

**Figure 5.** (a) Simulated number density as a function of aging time, along with the effective 2D number densities calculated from the experimental data (open orange circles). (b) Simulated precipitate area fraction as a function of aging time. (c) Simulated average Nd composition in the matrix as a function of aging time, compared with experimental APT measurements (filled orange circles, with error bars showing the statistical uncertainties). The inset shows a magnified view for 3–7 h. For all plots, solid green lines and shaded regions represent the mean and ±1 standard deviation (SD) of 3 independent simulations, respectively.

Representative snapshots of the microstructure from the simulation for which the number density result was presented in figure 4 are shown in figure 6 for aging times $t_0$–$t_8$. The plate-like morphologies of β1 precipitates with a habit plane of $\{1\bar{1}00\}_\alpha$ in the matrix is reproduced in simulations. All six precipitate growth directions, consistent with the three correspondence variants and their two rotation variants, are observed in the simulations, which agrees with experiments. This is notable since the model did not account for interfacial energy anisotropy; the results are solely due to anisotropic elastic properties.

The microstructural evolution shown in figure 6 exhibits different kinetic stages, supporting the trend in the number density evolution shown in figure 5. Nucleation of $\beta_1$ precipitates occurs gradually in the matrix in regions of higher solute supersaturation, with a time-dependent rate consistent with the incubation period. In the early stage (figure 6(a)–(c)), the nucleation rate is relatively low, as evidenced by the small number of nuclei observed. These nuclei experience significant growth along the primary direction during this period. Subsequently, a transition to a rapid nucleation regime occurs within figure 6(c)–(f), resulting in a significant increase in the number of observed nuclei. Beyond figure 6(g), the number of precipitates starts to decrease as some small precipitates disappear (see the ovals in figure 6(h)–(i)), which indicates that the system is in a coarsening regime. The simulations qualitatively reproduce these key features of the microstructure evolution, consistent with experimental results. A video showing the evolution of simulated number density and microstructure side by side is provided as figure S4 of the SI, offering a dynamic view of microstructural evolution during the simulation. Representative microstructure snapshots from all three independent simulations are presented in figure S5.

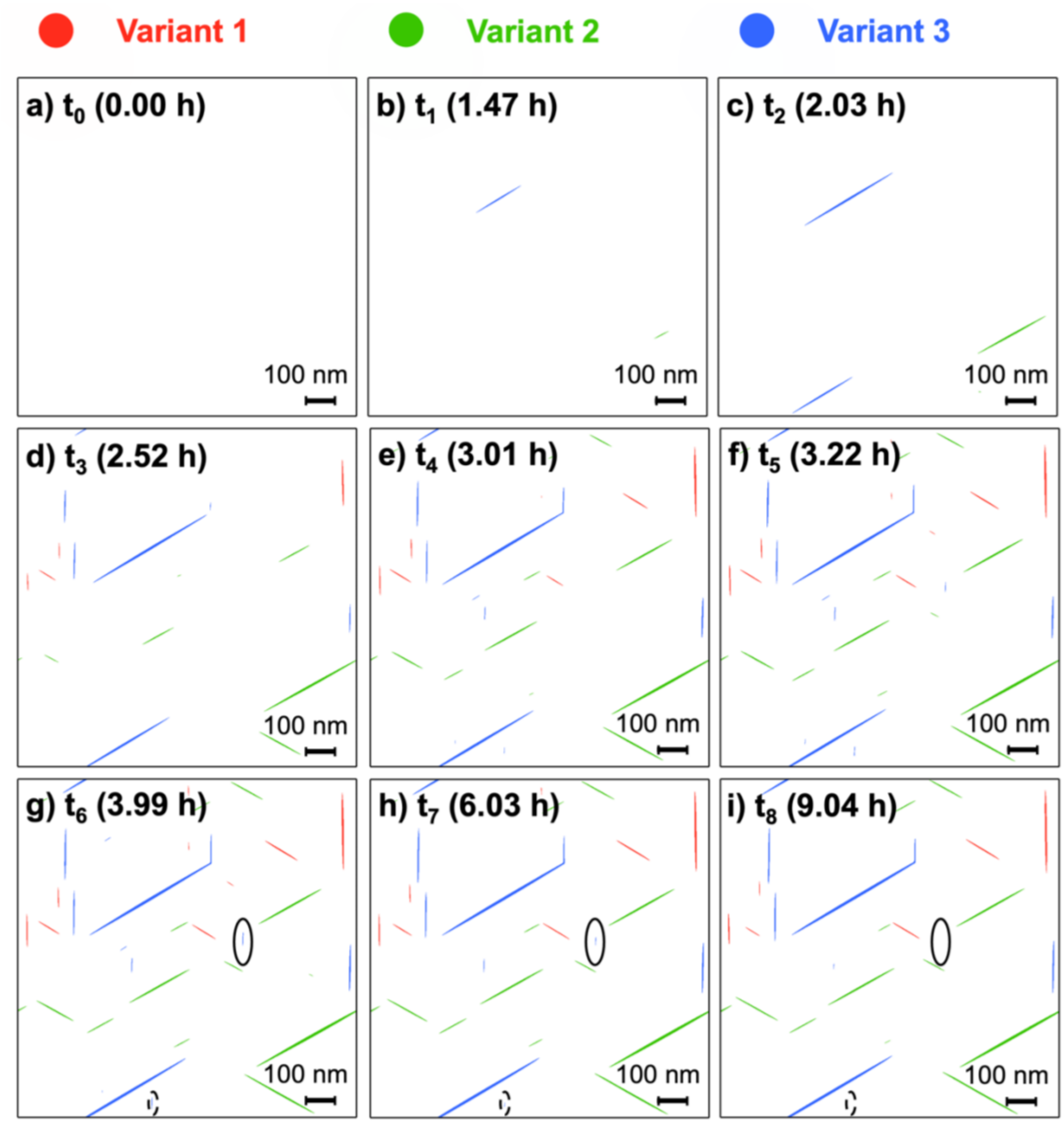

**Figure 6.** (a)–(i) Simulated microstructure snapshots at selected aging times, $t_0$–$t_8$, with different colors representing distinct correspondence variants of the $\beta_1$ precipitates. Solid black ovals and dashed black ovals in (g)–(i) highlight two small precipitates that diminish in size and eventually disappear during coarsening.

The growth of nuclei is strongly influenced by local solute diffusion fields. Figure 7 shows the matrix composition at (a) 2.31 h, (b) 3.01 h, and (c) 4.08 h for the same simulation. By 2.31 h in figure 7(a), significant spatial variations in the supersaturation arise due to the few precipitates that have formed early and grown large. Growth of precipitates may be suppressed depending on their environment, as solute may be depleted by other nearby precipitates. For example, two small precipitates having similar sizes at 2.31 h are highlighted in figure 7(a), one with a solid red oval and the other with a dashed red oval. By 3.01 h in figure 7(b), the precipitate marked with dashed oval grows significantly larger than the solid-oval one, which has more neighbors competing for solute, one of which formed early and has grown significantly. On the other hand, the precipitate marked with a dashed red oval formed away from other precipitates, and thus it is able to grow to much larger size. The overall supersaturation decreases as more nucleation events occur and precipitates grow, as shown in figure 7(a)–(c). Due to this depletion, neither of the two precipitates experiences significant growth from figure 7(b) to (c). By the latter time, nucleation ceases, growth slows, and the coarsening stage begins.

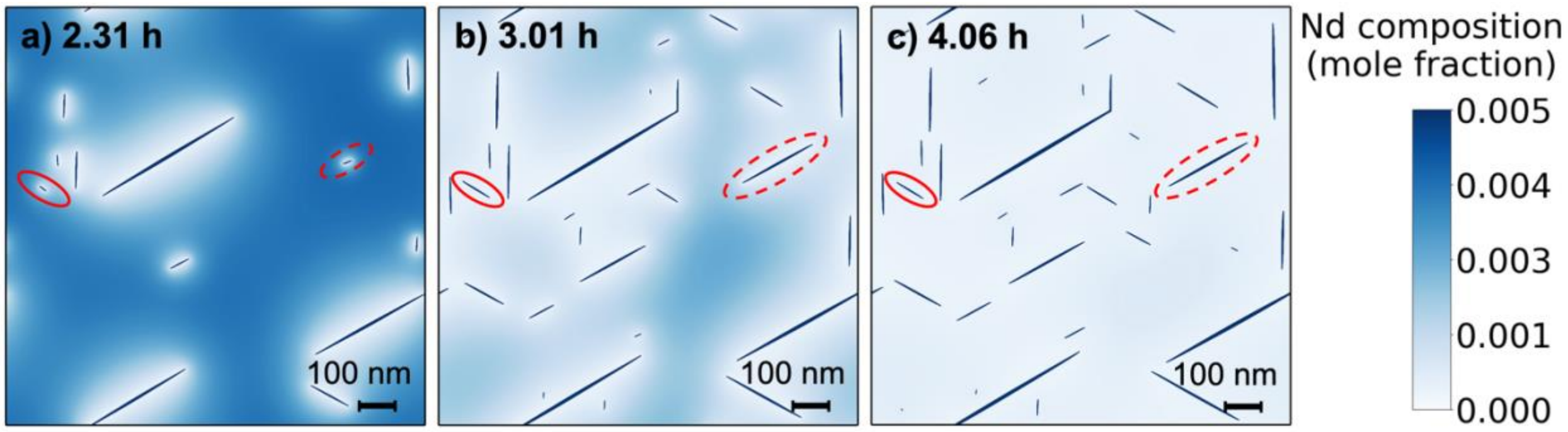

**Figure 7.** Nd composition at (a) 2.31 h, (b) 3.01 h and (c) 4.06 h. To better contrast the precipitates and the matrix, a maximum value of 0.005 was used in the colorbar, while the maximum possible matrix composition is 0.004. Composition values greater than 0.005 were truncated. A solid red oval and a dashed red oval in the plots highlight two precipitates of comparable size at 2.31 h. The 4.06 h snapshot shows the depletion of solute supersaturation in the matrix as nucleation and growth progress.

## 5. Conclusions

In this work, we developed and validated a workflow that combines experimental measurements and literature data to parameterize a 2D phase-field model for stochastic nucleation, growth, and coarsening of $\beta_1$ precipitates in Mg–Nd alloys. The workflow addresses the challenges associated with limited input data, especially in modeling the nucleation stage. The resulting parameterized model captures both the evolution of individual morphologies of the $\beta_1$ precipitates, as well as the overall evolution of the precipitate microstructure. Specifically, we examined the evolution of precipitate number density determined from experiments, in which it exhibits a rapid increase during the nucleation stage followed by a slow decrease during the coarsening stage. Although the APT measurements for the average Nd composition of the matrix were sparse, the simulation results were in excellent agreement with the available data.

The limitations of this study primarily arise from necessary modeling approximations. Most importantly, 2D simulations that use the effective 2D number densities derived from estimated 3D experimental data cannot fully capture the 3D nature of the diffusional interactions between precipitates, which may be important in determining microstructure evolution. This is likely the reason for the increased discrepancies between the predicted and measured precipitate number densities during coarsening. In addition, using the uniform-supersaturation model for the initial optimization of the parameters potentially overconstrains the parameter space for optimization, leading to a suboptimal parameter set. However, this approach circumvents the need to perform a full sensitivity analysis, which can greatly increase the computational cost due to the large number of phase-field simulations that would be required to fully explore the parameter space.

Our workflow that uses experimental data to obtain model parameters and applies the parameterized model to the evolution of microstructure provides a robust framework that can be (1) employed to optimize the aging time for desired microstructures and (2) extended to study precipitation phenomena in a wide range of alloy systems beyond the $\beta_1$ phase in Mg–Nd alloys and for different aging conditions. We believe our results can be improved by performing experimental measurements of both composition and precipitate number density with a higher time resolution, particularly during the nucleation stage. This will provide more data to fit to and will allow us to find the optimal model parameters with higher precision. In addition, the framework could benefit from incorporating a Bayesian method to efficiently explore the combined model parameter space. Finally, a combination of 3D simulations and 3D characterization will be required to fully understand the complex interplay between thermodynamics, interfacial and elastic anisotropies, and the kinetics of nucleation, growth, and coarsening in Mg–Nd and other Mg–RE alloys. The methodology presented here provides a foundation on which these extensions can be built.

## Acknowledgements

This work is supported by the U.S. Department of Energy, Office of Basic Energy Sciences, Division of Materials Sciences and Engineering, under Award No. DE-SC0008637 as part of the Center for Predictive Integrated Structural Materials Science (PRISMS Center) at the University of Michigan. This work used the allocation MSS160003 from the Advanced Cyberinfrastructure Coordination Ecosystem: Services & Support (ACCESS) program, which is supported by National Science Foundation grants Nos. 2138259, 2138286, 2138307, 2137603, and 2138296. This research also used resources of the National Energy Research Scientific Computing Center (NERSC), a Department of Energy User Facility, under NERSC award BES-ERCAP 0027826. The authors gratefully acknowledge Dr. Zhihua Huang and Prof. Amit Misra for sample preparation, providing TEM measurements and related data, as well as for helpful discussions that contributed to improving this work.

**Supplementary Information**

for

**Phase-field simulations of nucleation, growth, and coarsening of $\beta_1$ precipitates in Mg–Nd alloys**

L. Shi[1], S. DeWitt[1], D. Montiel[1], Q. Shi[1], J. Allison[1], and K. Thornton[1,2,*]

[1]Department of Materials Science and Engineering, University of Michigan, Ann Arbor, MI 48109, United States of America

[2]Department of Nuclear Engineering and Radiological Sciences, University of Michigan, Ann Arbor, MI 48109, United States of America

*Author to whom any correspondence should be addressed.

**Email**: kthorn@umich.edu

## S1. Calculation of 3D number density from precipitate number counts in TEM images

To demonstrate the applicability of equation (21) to circular platelets, we first derive the expression for the 3D number density assuming a single precipitate diameter, show that it is identical to the spherical-particle result, and then consider the effect of the size distribution.

The TEM sample can be modeled as a planar foil of thickness, $t_{foil}$, and area, $A_{foil}$. Assuming that this foil is parallel to the basal plane of Mg, the $\beta_1$ precipitates can be approximated as circular plates oriented in directions perpendicular to the foil, as shown in figure S1. We assume that $N$ plate-shaped particles are randomly distributed in space across a sufficiently large representative volume, $V$. For plates of equal diameter, $d_p$, and thickness, $\lambda_p$, we can relate the 3D number density, $N_V$, defined as the number of particles per volume, to a 2D number density, $N_A$, defined as $N'/A_{foil}$, where $N'$ is the number of particles that intersect the foil (the same number that is detected in the TEM). The 3D number density can be written as

$$N_V = \frac{N}{V} = \frac{N'}{V'}, \tag{S1}$$

where $V'$ is the volume that encloses all the particles that intersect with the foil. Because the plate-shape particles are oriented perpendicularly to the foil, all particles centered within a distance of $(d_p + t_{foil})/2$ from the middle of the foil will intersect. Thus, we can obtain $V'$as

$$V' = A_{foil}(d_p + t_{foil}) \tag{S2}$$

Substituting $V'$ in equation (S1), we obtain

$$N_V = \frac{N'}{A_{foil}(d_p + t_{foil})} = \frac{N_A}{(d_p + t_{foil})}. \tag{S3}$$

This equation is identical to that for equally sized spheres derived in Ref. [1], where $d_p$ is the diameter of the spherical particles. Therefore, from geometric considerations, it is clear that the analysis for spheres can also be applied to circular plates oriented in a direction perpendicular to the foil.

Having demonstrated the applicability of the spherical particle analysis, we now consider the limitations of equation (S3). The first limitation is that the derivation assumes particles of equal size. Second, it requires knowledge of the actual particle diameter, which is not readily available from a TEM image unless the film thickness is comparable to or greater than the particle size. Several expressions to estimate $N_V$ from measured diameters for a distribution of particle sizes have been obtained [2, 3], but we did not adopt these since we do not have the size distribution of the precipitates. We instead employ the following expression proposed by Sonderegger [4], which was demonstrated to be accurate even when the particle size distribution is unknown:

$$N_V = \frac{1}{A_{foil}} \sum_{i=1}^{N'} \frac{1}{t_{foil} + d_{m,i}}, \tag{S4}$$

where the sum is over all precipitates intersecting the foil and $d_{m,i}$ is the measured projected diameter of the $i^{\text{th}}$ precipitate. Qualitatively, this expression can be obtained by defining a partial number density for each particle based on equation (S3),

$$N_{V,i} = \frac{N_i'}{A_{foil}(d_{m,i} + t_{foil})}, \tag{S5}$$

and then summing over all particles, assuming the particle diameter can be approximated by the measured diameter.

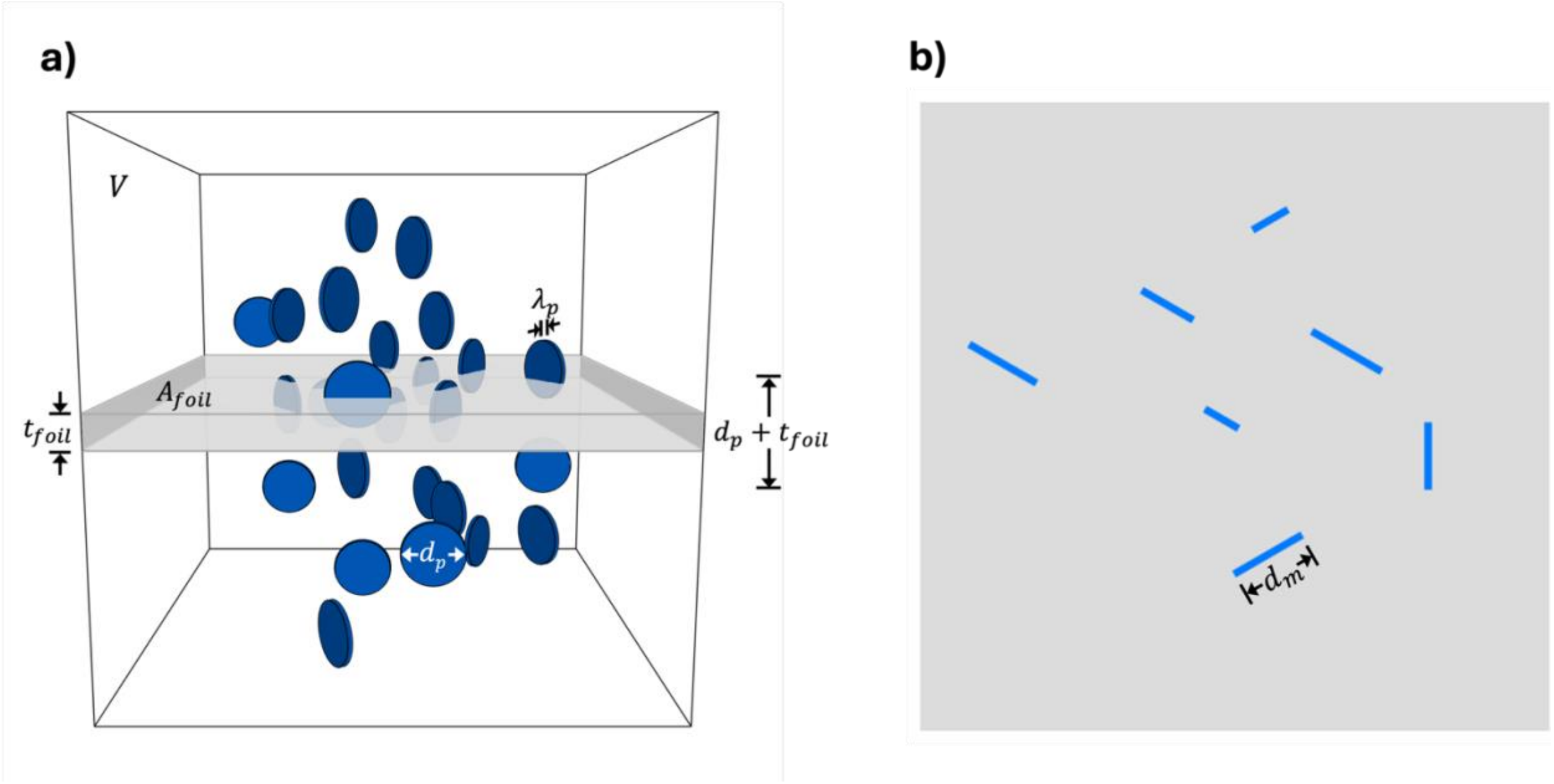

**Figure S1.** (a) 3D side view of a representative volume in which plate-shaped particles are randomly distributed in space. The foil is represented by the gray slab in the middle. (b) Top view of the volume. The elongated rectangles represent the sections of precipitates with projected diameter, $d_m$, that intersect the foil.

## S2. Model parameters

**Table S1.** Summary of model parameters in physical and dimensionless values.

| Parameter and dimensional symbol | Physical value with unit | Dimensionless value |
|---|---|---|
| Characteristic energy density $f^*$ | $2 \times 10^9$ J m$^{-3}$ | 1 |
| Characteristic length $l^*$ | $1 \times 10^{-9}$ m | 1 |
| Characteristic time $t^*$ | $l^{*2}/D$ | 1 |
| Simulation domain size | $(1.1\times10^{-6}\ \text{m})^2$ | $1100^2$ |
| Time step $\Delta t$ | $0.0025t^*$ | 0.0025 |
| Nucleation attempt time interval $\Delta t_{nuc}$ | $2.5t^*$ | 2.5 |
| Nucleus semi-major axis | $3\times10^{-9}$ m | 3 |
| Nucleus semi-minor axis | $2\times10^{-9}$ m | 2 |
| Nucleus holding time | $100t^*$ | 100 |
| Diffusion coefficient $D$ | $(1.1\times10^{-6}\ \text{m})^2/t^*$ | 1 |
| Kinetic interface mobility $L$ | $(8\times10^{9}\ \text{J m}^{-3})/t^*$ | 4 |
| Gradient energy coefficient $K$ | $1.023\times10^{26}$ J m$^{-3}$ s$^{-1}$ | $5.115\times10^{-2}$ |
| Energy barrier constant $W$ | $4.4\times10^{8}$ J m$^{-3}$ | 0.22 |
| SFTS tensors $\epsilon^{\eta_1}$, $\epsilon^{\eta_2}$, $\epsilon^{\eta_3}$ | $\epsilon^{\eta_1}=\begin{bmatrix}0.1111 & 0.0962 & 0\\ 0.0962 & 0 & 0\\ 0 & 0 & 0\end{bmatrix}$ $\epsilon^{\eta_2}=\begin{bmatrix}-0.0556 & 0 & 0\\ 0 & 0.1666 & 0\\ 0 & 0 & 0\end{bmatrix}$ $\epsilon^{\eta_3}=\begin{bmatrix}0.1111 & -0.0962 & 0\\ -0.0962 & 0 & 0\\ 0 & 0 & 0\end{bmatrix}$ | $\begin{bmatrix}0.1111 & 0.0962 & 0\\ 0.0962 & 0 & 0\\ 0 & 0 & 0\end{bmatrix}$ $\begin{bmatrix}-0.0556 & 0 & 0\\ 0 & 0.1666 & 0\\ 0 & 0 & 0\end{bmatrix}$ $\begin{bmatrix}0.1111 & -0.0962 & 0\\ -0.0962 & 0 & 0\\ 0 & 0 & 0\end{bmatrix}$ |
| Elastic constants $C_{ijkl}$ | $C_{1111}=C_{2222}=62.6$ GPa<br>$C_{3333}=64.9$ GPa<br>$C_{1122}=26.0$ GPa<br>$C_{1133}=C_{2233}=20.9$ GPa | 31.3<br>32.45<br>13.0<br>10.45 |
| | $C_{2323} = C_{1313} = 13.3$ GPa | 6.65 |
| | $C_{1212} = 18.3$ GPa | 9.15 |
| Chemical free energy density coefficients $A_0$, $A_1$, $A_2$, $B_0$, $B_1$, $B_2$ | $A_0 = 4.00 \times 10^{-1}$ J mol$^{-1}$ | $1.43 \times 10^{-5}$ |
| | $A_1 = -4.23 \times 10^{3}$ J mol$^{-1}$ | $-1.51 \times 10^{-1}$ |
| | $A_2 = 1.12 \times 10^{7}$ J mol$^{-1}$ | $4.00 \times 10^{2}$ |
| | $B_0 = 1.71 \times 10^{4}$ J mol$^{-1}$ | $6.11 \times 10^{-1}$ |
| | $B_1 = -1.41 \times 10^{5}$ J mol$^{-1}$ | $-5.04$ |
| | $B_2 = 2.89 \times 10^{5}$ J mol$^{-1}$ | $1.03 \times 10^{1}$ |
| Nucleation rate parameters $\rho_1$, $\rho_2$, $\tau_{in}$ | $\rho_1 = (1.70 \times 10^{9}\ \mathrm{m}^{-2})/t^*$ | $1.70 \times 10^{-9}$ |
| | $\rho_2 = 1.74 \times 10^{-7}$ | $1.73 \times 10^{-7}$ |
| | $\tau_{in} = 8.08 \times 4t^*$ | $8.08 \times 4$ |

## S3. Approximate expression for $\Delta G_v$

Consider the transformation from a solid-solution matrix phase, α, to a nucleus, $\beta_1$ in a small region. Following Ref. [5], upon removal of a small amount of material with the nucleus composition, $x_{Nd}^{\beta_1}$, from the α phase, the total free energy of the system per mole removed decreases by

$$\Delta G_m^1 = \mu_{Mg}^{\alpha} x_{Mg}^{\beta_1} + \mu_{Nd}^{\alpha} x_{Nd}^{\beta_1}, \tag{S6}$$

where $x_i^j$ is the mole fraction of species $i$ in phase $j$, and $\mu_i^j$ is the chemical potential of species $i$ in phase $j$, defined as

$$\mu_i^j = \left(\frac{\partial G^j}{\partial N_i}\right)_{T,P,N_{k\neq i}}. \tag{S7}$$

The subscript $m$ indicates that the quantity is per mole. In equation (S7), $G^j$ is the contribution from phase $j$ to the Gibbs free energy of the system, and $N_i$ is the number of moles of species $i$. If the removed material transforms into the $\beta_1$ phase, the free energy of the system per mole added increases by

$$\Delta G_m^2 = \mu_{Mg}^{\beta_1} x_{Mg}^{\beta_1} + \mu_{Nd}^{\beta_1} x_{Nd}^{\beta_1}. \tag{S8}$$

Therefore, the driving force for nucleation is expressed as

$$\Delta G_m^{nuc} = \Delta G_m^2 - \Delta G_m^1$$
$$= \left(\mu_{Mg}^{\beta_1} - \mu_{Mg}^{\alpha}\right) + \left[\left(\mu_{Nd}^{\beta_1} - \mu_{Nd}^{\alpha}\right) - \left(\mu_{Mg}^{\beta_1} - \mu_{Mg}^{\alpha}\right)\right] x_{Nd}^{\beta_1}. \tag{S9}$$

The change of free energy per unit volume, $\Delta G_v$, can be obtained from $\Delta G_m^{nuc}$ as

$$\Delta G_v = \frac{\Delta G_m^{nuc}}{V_m}, \tag{S10}$$

where $V_m$ is the molar volume. The subscript $v$ indicates that the quantity is per volume.

Considering that the transformation occurs within a constant volume and the pressure change is negligible, we can assume that the change in the Helmholtz free energy is approximately equal to that of the Gibbs free energy. We also assume that the nucleus composition, $x_{Nd}^{\beta_1}$, can be approximated by the equilibrium composition in the $\beta_1$ phase, $x_e^{\beta_1}$. As in Ref. [6], parabolic free energy densities (equations (3) and (4)) of each phase are used, which determine the chemical potentials in equation (S9):

$$\mu_{Mg}^{\alpha} = f_\alpha(x_0) - x_0 \left.\frac{\partial f_\alpha(x_{Nd}^{\alpha})}{\partial x_{Nd}^{\alpha}}\right|_{x_{Nd}^{\alpha}=x_0} = -A_2(x_0)^2 + A_0; \tag{S11}$$

$$\mu_{Nd}^{\alpha} = f_\alpha(x_0) + (1 - x_0) \left.\frac{\partial f_\alpha(x_{Nd}^{\alpha})}{\partial x_{Nd}^{\alpha}}\right|_{x_{Nd}^{\alpha}=x_0}$$
$$= A_2[-(x_0)^2 + 2x_0] + A_1 + A_0; \tag{S12}$$

$$\mu_{Mg}^{\beta_1} = f_\alpha(x_e^{\alpha}) - x_e^{\alpha} \left.\frac{\partial f_\alpha(x_{Nd}^{\alpha})}{\partial x_{Nd}^{\alpha}}\right|_{x_{Nd}^{\alpha}=x_e^{\alpha}} = -A_2(x_e^{\alpha})^2 + A_0; \tag{S13}$$

$$\mu_{Nd}^{\beta_1} = f_\alpha(x_e^{\alpha}) + (1 - x_e^{\alpha}) \left.\frac{\partial f_\alpha(x_{Nd}^{\alpha})}{\partial x_{Nd}^{\alpha}}\right|_{x_{Nd}^{\alpha}=x_e^{\alpha}}$$
$$= A_2[-(x_e^{\alpha})^2 + 2x_e^{\alpha}] + A_1 + A_0. \tag{S14}$$

Here, $x_e^{\alpha}$ is the equilibrium composition in the matrix phase, and $x_0$ is the initial solute composition in the solid solution.

Substituting expressions (S11)–(S14) into (S9) yields

$$\Delta G_m^{nuc} = \left(\mu_{Mg}^{\beta_1} - \mu_{Mg}^{\alpha}\right) + \left[\left(\mu_{Nd}^{\beta_1} - \mu_{Nd}^{\alpha}\right) - \left(\mu_{Mg}^{\beta_1} - \mu_{Mg}^{\alpha}\right)\right] x_{Nd}^{\beta_1}$$
$$= 2A_2\left(x_e^{\beta_1} - x_e^{\alpha}\right)\left[\frac{\Delta x}{2\left(x_e^{\beta_1} - x_e^{\alpha}\right)} - 1\right]\Delta x, \tag{S15}$$

where $\Delta x = x_0 - x_e^\alpha$ is the supersaturation. In the Mg–Nd system studied, $\frac{\Delta x}{2\left(x_e^{\beta_1} - x_e^\alpha\right)} \ll 1$ for the entire range of possible supersaturation values during aging. Thus, we approximate $\Delta G_v$ as proportional to $\Delta x$

$$\Delta G_v \approx -\frac{2A_2\left(x_e^{\beta_1} - x_e^\alpha\right)}{V_m}\Delta x. \tag{S16}$$

## S4. Simplified expression for $\Delta G^*$ and $\rho_2$

We start from the expression for the nucleation rate from classical nucleation theory (equation (16) in the main text):

$$J = ZN_n\beta \exp\left(-\frac{\Delta G^*}{kT}\right) \exp\left(-\frac{\tau_{in}}{t}\right), \tag{S17}$$

where

$$\Delta G^* = \frac{16\pi\gamma^3}{3(|\Delta G_v| - |\Delta G_{strain}|)^2}. \tag{S18}$$

Neglecting the term $\Delta G_{strain}$, we obtain

$$\Delta G^* \simeq \frac{16\pi\gamma^3}{3{\Delta G_v}^2}. \tag{S19}$$

The effect of this approximation is discussed in section S5. Here, $\Delta G_v$ could be directly taken from CALPHAD free energy data. However, it is useful to have an analytical expression for $\rho_2$ to determine its fitting bounds. Substituting equation (S16), based on the quadratic free energy, into equation (S19), we obtain

$$\Delta G^* \simeq \frac{4\pi\gamma^3 V_m^2}{3(A_2)^2\left(x_e^{\beta_1} - x_e^\alpha\right)^2 \Delta x^2}. \tag{S20}$$

Finally, substituting $\Delta G^*$ into equation (S17), we obtain

$$J(\boldsymbol{r}, t) = ZN_n\beta \exp\left(-\frac{\rho_2}{\Delta x^2}\right) \exp\left(-\frac{\tau_{in}}{t}\right), \tag{S21}$$

with $\rho_2$ given by

$$\rho_2 = \frac{4\pi\gamma^3 {V_m}^2}{3kT(A_2)^2\left(x_e^{\beta_1} - x_e^\alpha\right)^2}, \tag{S22}$$

which is used to set the bounds for $\rho_2$ as described in section 4.4. Since we treat $\rho_2$ as an effective parameter that is adjusted to reproduce the experimental precipitate number density evolution, the quadratic approximations of the free energies do not affect the result.

## S5. The effect of $\Delta G_{strain}$ on the nucleation rate $J$

Assuming the system is initially unstrained, the change in strain energy of the system per volume of the nucleus upon its introduction, $\Delta G_{strain}$, is

$$\Delta G_{strain} = \frac{1}{V_{nucleus}} \int_V f_{el} dV, \tag{S23}$$

where $V_{nucleus}$ is the volume of a single nucleus introduced to the Mg–Nd system and $f_{el}$ is the elastic energy density that is introduced by the single nucleus across the domain volume, $V$. For our 2D phase-field simulations, we have

$$\Delta G_{strain} = \frac{1}{A_{nucleus}} \int_A f_{el} dA, \tag{S24}$$

where $A_{nucleus}$ is the area of the nucleus and $A$ denotes the domain area. For the values of $C_{ijkl}$ and $\epsilon_{ij}^0$ reported in section 4.3, we obtained a value of $\Delta G_{strain} = 1.33 \times 10^8$ J m$^{-3}$ from a 2D phase-field simulation for a seed size similar to the critical nucleus size.

Using the values of $A_2$, $x_e^{\beta_1}$, and $x_e^{\alpha}$ reported in section 4.3, along with $\Delta x = 3.70 \times 10^{-3}$ (which corresponds to the initial supersaturation in the system), we calculate $\Delta G_v = -1.44 \times 10^9$ J m$^{-3}$. By comparing with the strain energy term, we obtain the ratio $|\Delta G_{strain}|/|\Delta G_v| \simeq 0.092 < 1/10$; therefore, neglecting the strain energy is a reasonable assumption during the early stage of nucleation. The last few seeded nuclei in the simulations occur in regions where $\Delta x$ is around 0.001, leading to $\Delta G_v \approx -3.6 \times 10^8$ J m$^{-3}$. Although this value is still smaller in magnitude than $\Delta G_{strain}$, it yields the ratio $|\Delta G_{strain}|/|\Delta G_v| \simeq 0.37$, which indicates that strain energy may affect the nucleation rate toward the end of the simulation.

To estimate the effect of $\Delta G_{strain}$ on the nucleation rate, we compared $J_0$, the nucleation rate used in the simulation, and $J_{strain}$, which includes $\Delta G_{strain} = 1.33 \times 10^8$ J m$^{-3}$ in otherwise the same expression and with the same parameter values (see table 4 and equation (28)). The $J_{strain}$ expression is obtained from equations (S16)–(S18):

$$J_{strain} = \rho_1 \exp\left[-\rho_2\left(\Delta x - \left(\frac{3kT\rho_2}{16\pi\gamma^3}\right)^{\frac{1}{2}}\Delta G_{strain}\right)^{-2}\right]\exp\left(-\frac{\tau_{in}}{t}\right). \quad \text{(S25)}$$

The calculated $J_0$ and $J_{strain}$ are shown in figure S2. The small difference between $J_0$ (the solid blue curve) and $J_{strain}$ (the solid red curve) indicates that the effect of $\Delta G_{strain}$ on the nucleation rate is not significant in our simulations. Specifically, the relative difference in the maximum number density yielded by $J_{strain}$ is only 6.7% with respect to $J_0$.

Furthermore, we performed additional analysis to demonstrate that a similar nucleation rate (as a function of time) to $J_0$ can be obtained by tuning the nucleation rate parameters $\rho_1$, $\rho_2$ and $\tau_{in}$ when including $\Delta G_{strain}$. The resulting nucleation rate, $J_{strain}^{modified}$, had $\rho_2 = 1.09 \times 10^{-7}$ and $\tau_{in}$= 3.05 h (vs. the original $1.74 \times 10^{-7}$ and 3.00 h, respectively); the $\rho_1$ value remained at $4.611 \times 10^{13}\ \mathrm{m^{-2}\ h^{-1}}$. This implies that the impact of $\Delta G_{strain}$ on the nucleation rate can be largely absorbed by tuning the nucleation rate parameters. Since our nucleation rate parameters used in the uniform-supersaturation model were obtained through fitting to experiments, we believe that the effect of $\Delta G_{strain}$ is implicitly accounted for within these effective parameters.

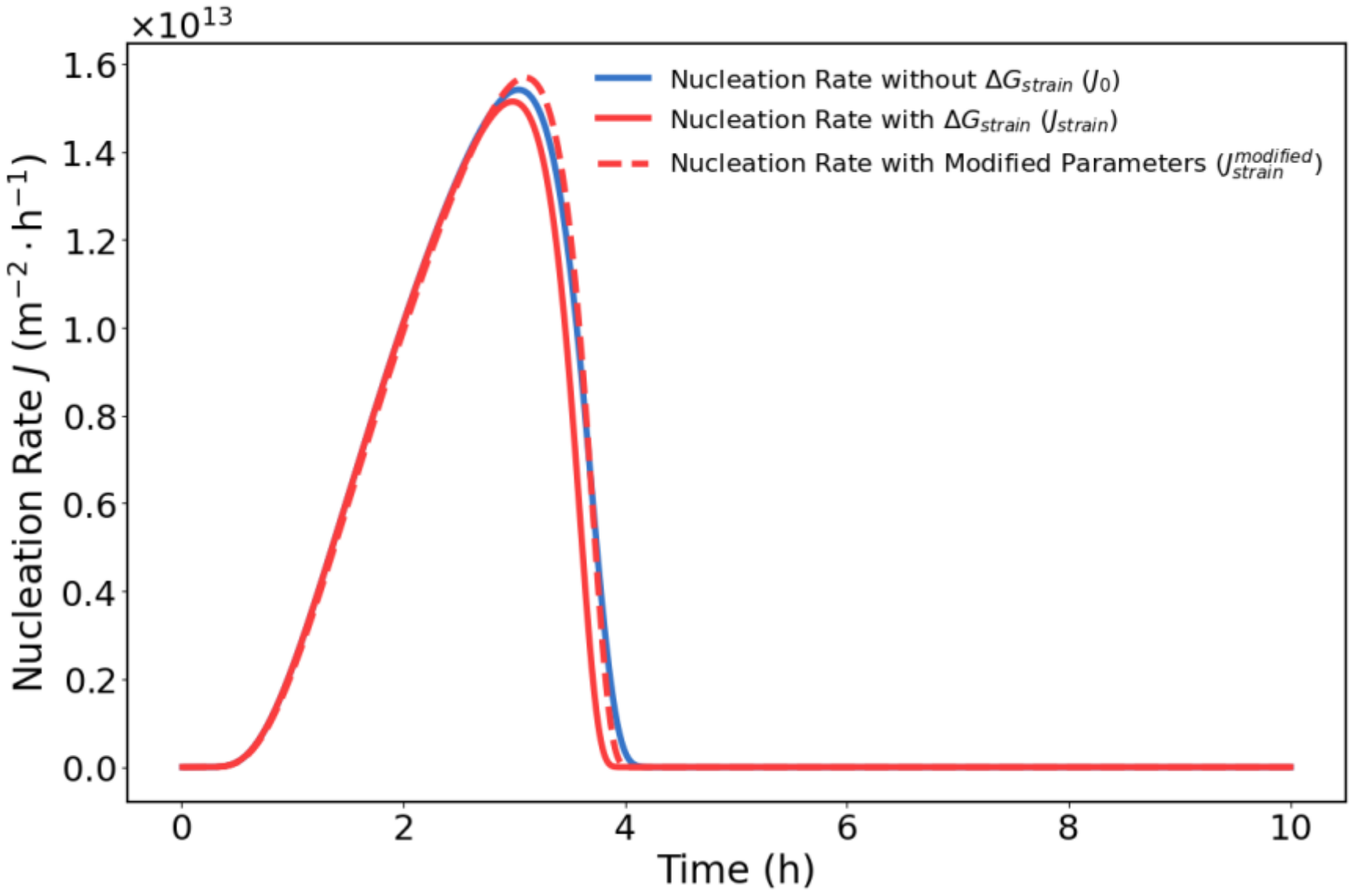

**Figure S2.** Nucleation rates as a function of time. The solid blue curve represents the nucleation rate without $\Delta G_{strain}$ ($J_0$), the solid red curve represents the nucleation rate incorporating $\Delta G_{strain}$ ($J_{strain}$), and the dashed red line shows the rate $J_s^{modified}$ after modifying the nucleation rate parameters in $J_{strain}$ to minimize the difference with respect to $J_0$.

## S6. Composition fit for initial estimation of the nucleation rate parameters

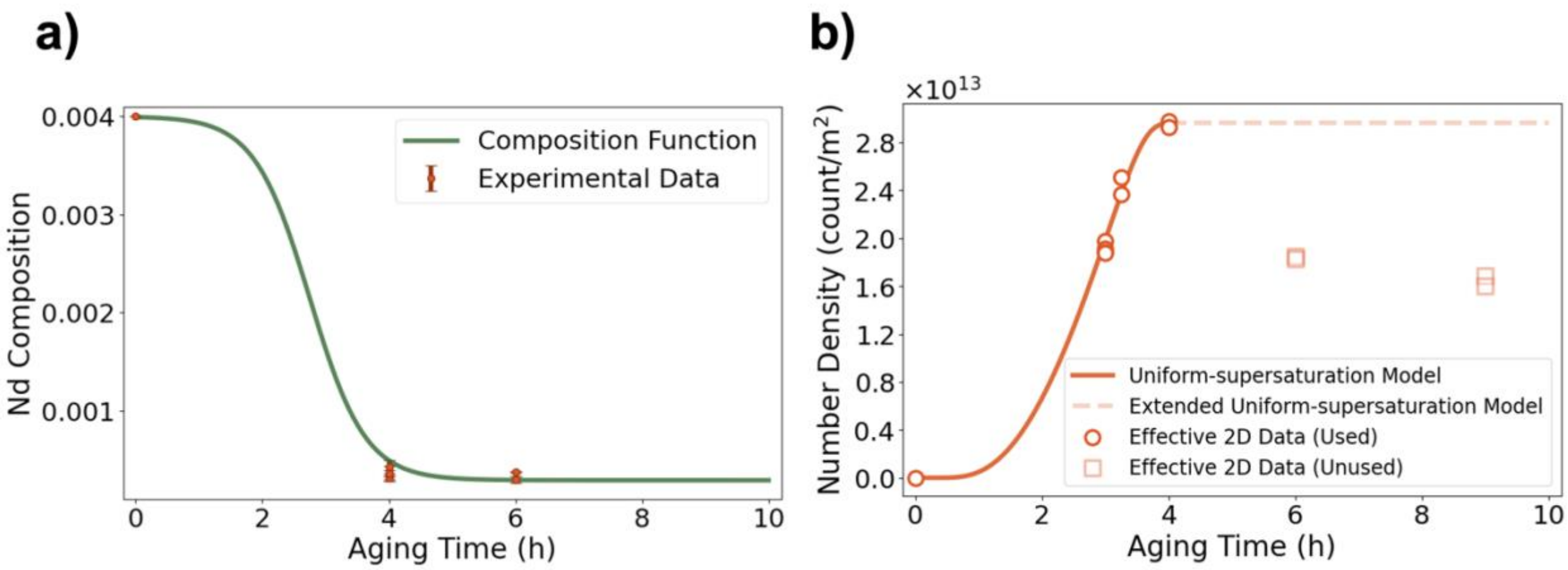

**Figure S3.** Fitted average matrix composition as a function of aging time, shown with experimental matrix composition data points.

## S7. Evolution over time for fully parameterized model

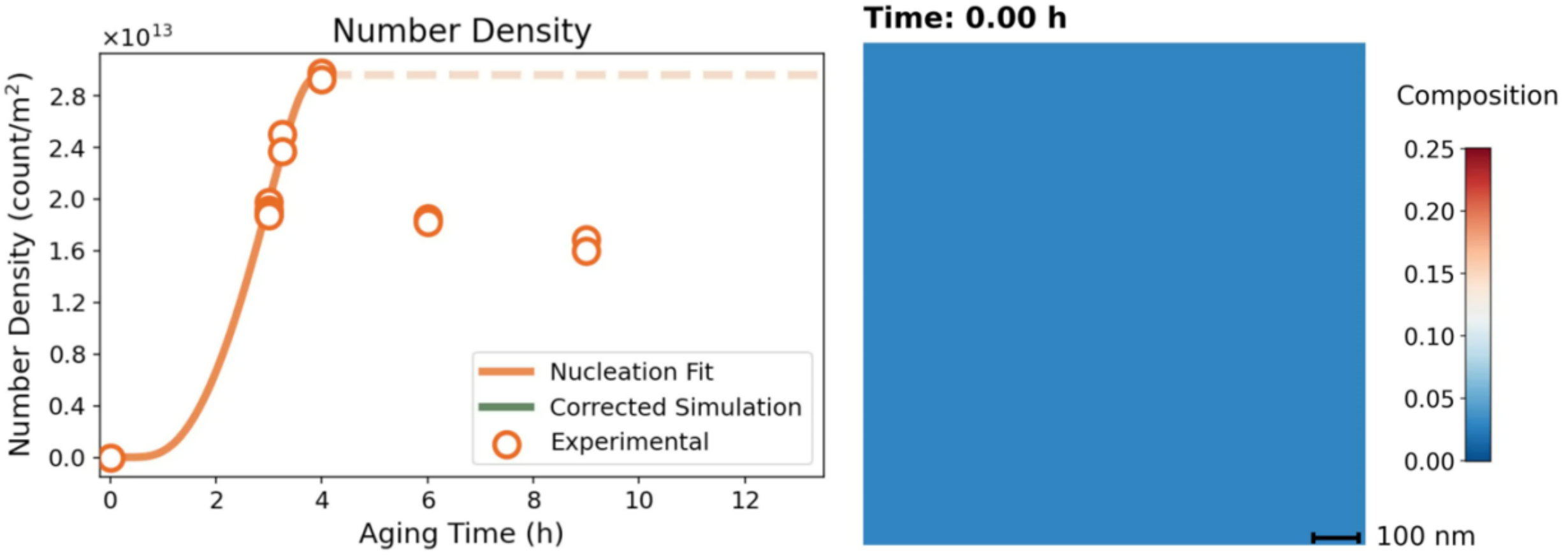

**Figure S4.** Video of simulated precipitate number density (left) and microstructure evolution (right) over time.

## S8. Microstructures from different simulations

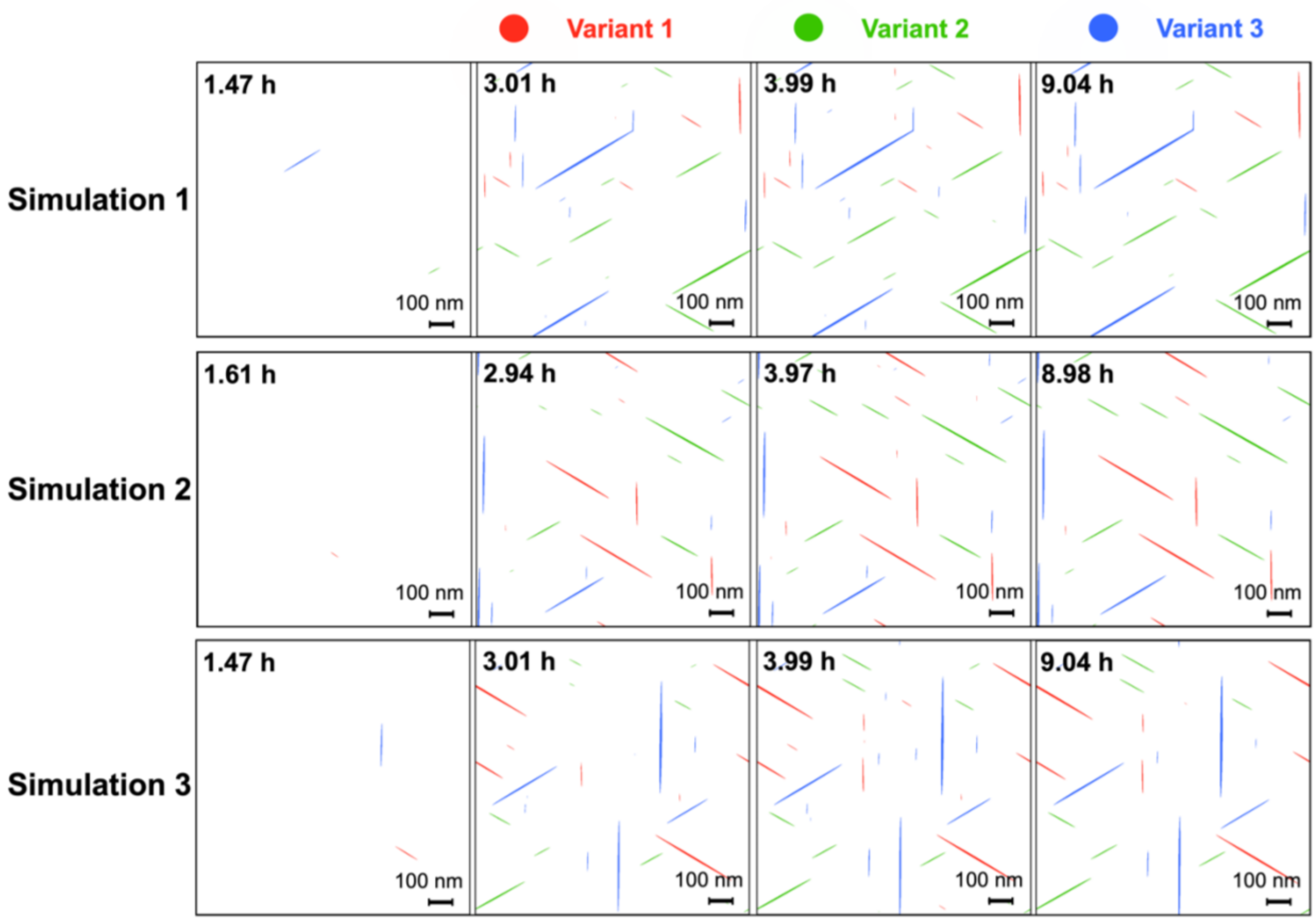

**Figure S5.** Simulated microstructure snapshots from three independent simulations at selected aging times, with different colors representing distinct precipitate variants of the $\beta_1$ precipitates. In Simulation 2, no nucleus was seeded before 1.61 h, while Simulations 1 and 3 experienced earlier nucleation.